\documentclass{jpp}
\usepackage{graphicx}
\usepackage{bm}
\usepackage{xcolor}

\usepackage[utf8]{inputenc}
\usepackage[T1]{fontenc}
\usepackage{amsmath}
\usepackage{subcaption}
\usepackage[normalem]{ulem}

\shorttitle{Shape gradient of stellarator coil complexity with respect to the plasma boundary}
\shortauthor{A. Carlton-Jones, E. J. Paul, and W. Dorland}

\title{Computing the shape gradient of stellarator coil complexity with respect to the plasma boundary}

\author{Arthur Carlton-Jones\aff{1},
  Elizabeth J. Paul\aff{2},
 \and William Dorland\aff{1}}

\affiliation{\aff{1}University of Maryland,
College Park, MD 20742, USA
\aff{2}Department of Astrophysical Sciences, Princeton University,
Princeton, NJ 08544, USA}

\begin{document}

\maketitle

\begin{abstract}
Coil complexity is a critical consideration in stellarator design. The traditional two-step optimization approach, in which the plasma boundary is optimized for physics properties and the coils are subsequently optimized to be consistent with this boundary, can result in plasma shapes which cannot be produced with sufficiently simple coils. To address this challenge, we propose a method to incorporate considerations of coil complexity in the optimization of the plasma boundary. Coil complexity metrics are computed from the current potential solution obtained with the REGCOIL code \citep{landreman2017}. While such metrics have previously been included in derivative-free fixed-boundary optimization \citep{drevlak}, we compute the local sensitivity of these metrics with respect to perturbations of the plasma boundary using the shape gradient \citep{Landreman2018}. We extend REGCOIL to compute derivatives of these metrics with respect to parameters describing the plasma boundary. In keeping with previous research on winding surface optimization \citep{Paul2018}, the shape derivatives are computed with a discrete adjoint method. In contrast with the previous work, derivatives are computed with respect to the plasma surface parameters rather than the winding surface parameters. To further reduce the resolution required to compute the shape gradient, we present a more efficient representation of the plasma surface which uses a single Fourier series to describe the radial distance from a coordinate axis and a spectrally condensed poloidal angle. This representation is advantageous over the standard cylindrical representation used in the VMEC code \citep{hirshman}, as it provides a uniquely defined poloidal angle, eliminating a null space in the optimization of the plasma surface. In comparison with previous spectral condensation methods \citep{Hirshman1998}, the modified poloidal angle is obtained algebraically rather than through the solution of a nonlinear optimization problem. The resulting shape gradient highlights features of the plasma boundary that are consistent with simple coils and can be used to couple coil and fixed-boundary optimization.
% , as it only requires a cosine series for a single variable rather than a sine and cosine series for two separate variables. It also 

% Coil complexity metrics can be computed with REGCOIL \citep{landreman2017}, which optimizes coil shapes using a linear least squares method based on a current potential approximation.  An adjoint method is used to compute these derivatives analytically. With this method, it is only necessary to solve the linear least-squares system used in REGCOIL twice, rather that for every surface parameter. This provides a great computational advantage over finite-difference differentiation and reduction of noise. Shape gradients of the REGCOIL metrics are computed from these derivatives; they tell us how these metrics can be altered through normal perturbations of the plasma surface. We also compute these derivatives and shape gradients while constraining one of the coil complexity metrics of REGCOIL to be fixed. This gives information on how to alter the plasma surface to be better reproduced by coils with a desired complexity. 
\end{abstract}

\section{Introduction}

Historically, several stellarators have been optimized with a two-staged approach, including W7-X \citep{Beidler1990}, NCSX \citep{Zarnstorff2001}, and HSX \citep{Anderson1995}. In the first stage, the outer boundary of the plasma, denoted by $S_{\text{plasma}}$, is optimized based on physical properties of the magnetohydrodynamic (MHD) equilibrium. This task is performed with optimization codes such as STELLOPT \citep{Spong2001} and ROSE \citep{drevlak}. During the second stage, the currents in the vacuum region must be chosen to be consistent with the boundary obtained from the first stage. This is typically formulated as obtaining the currents in the vacuum region, $\bm{J}^{\text{coil}}$, given the magnetic field due to the plasma currents, $\bm{B}^{\text{plasma}}$, such that the total normal magnetic field on $S_{\text{plasma}}$ vanishes,
\begin{align}
\bm{B}^{\text{plasma}}(\bm{r}) \cdot \hat{\bm{n}}(\bm{r}) + \frac{\mu_0}{4\pi} \int_{
\Omega_c 
} \frac{\bm{J}^{\text{coil}}(\bm{r}') \times (\bm{r}-\bm{r}') \cdot \hat{\bm{n}}(\bm{r})}{\rvert{\bm{r}-\bm{r}'\rvert^3}}  \, dV' = 0,
\label{eq:coil_design}
\end{align}
where $\Omega_c$ is the vacuum region. This is an integral equation of the first kind, which is known to be ill-posed in the sense that a unique solution may not exist and a small change in the prescribed data, $\bm{B}^{\text{plasma}} \cdot \hat{\bm{n}}$, may result in a large change in the solution, $\bm{J}^{\text{coil}}$ \citep{Kress1989,Imbert2019}. Given the ill-posedness, the coils problem has been reformulated as a regularized optimization problem. Under the approximation that the coils are filamentary curves, this nonlinear optimization problem is solved with the FOCUS \citep{Zhu2018} and ONSET \citep{Drevlak1998} codes. If $\bm{J}^{\text{coil}}$ is assumed to be a continuous surface current supported on a toroidal winding surface, $S_{\text{coil}}$, a convex optimization problem is obtained, which is solved with the REGCOIL \citep{landreman2017} and NESCOIL \citep{merkel1987solution} codes. We refer to this technique as a current potential formulation, as the divergence-free current density can be expressed as the gradient of a potential function. Given the inherent ill-posedness in the coil design problem \eqref{eq:coil_design}, these coil optimization problems can be formulated to favor simple coil shapes that can be constructed at a lower cost. 

Decoupling the physics and practical considerations of a design in this way is thought to have several possible advantages. By initially ignoring engineering considerations, one may obtain a configuration with enhanced confinement properties. This approach may also enable one to explore a wider space of coil designs, allowing for multiple coil topologies or the incorporation of permanent magnets. A major shortcoming of this approach is the possibility of arriving at a configuration that cannot be produced with sufficiently simple coils or magnets. 
% As the coil design problem \eqref{eq:coil_design} is ill-posed, small changes in the plasma equilibrium can result in large changes in the coil shapes. 
% Due to the ill-posedness of the coil design problem, a minor improvement in confinement properties may give rise to an untolerable increase in coil complexity.
% As was stated in the 2018 report of the National Stellarator Coordinating Committee \citep{Gates2018}, ``The highest priority for technology is to better integrate the engineering design with the physics design at the earliest possible stage." 
For this reason, it is favorable to integrate engineering considerations of the coils with stellarator equilibrium optimization.

This priority can be addressed with several techniques. One option is to directly optimize coils based on free-boundary solutions of the MHD equilibrium equations, eliminating the need to design coils as a second step. This approach was used in the final stages of the NCSX design \citep{Hudson2002,Strickler2003}, and recent developments have enabled more efficient direct optimization of coils with adjoint methods \citep{Giuliani2020}. This direct optimization approach is sometimes more challenging for several reasons. Free-boundary equilibrium calculations tend to be more expensive than fixed-boundary calculations, as they often require iterations between an equilibrium solve and vacuum field calculations. This iterative scheme will not always converge in practice, hence the historical use of the more robust fixed-boundary method. A novel proposed algorithm may reduce the cost associated with the free-boundary solve by eliminating the need for such iterations \citep{Henneberg2020}. 

There are several alternatives to free-boundary optimization. An algorithm has been proposed \citep{hudson} to simultaneously optimize the plasma boundary and a set of filamentary coils by minimizing the coil complexity at fixed confinement properties. Another approach is to incorporate metrics of coil complexity into the fixed-boundary equilibrium design. To avoid the simultaneous nonlinear optimization of filamentary coils, a current potential model can be used. We note that the complexity of the current potential solution does not necessarily correlate with the complexity of filamentary coils. For example, the normal field is typically reduced when the winding surface is close to the plasma surface in a current potential model, but it is increased when filamentary coils are close to the plasma boundary due to coil ripple. Nonetheless, the current potential model provides a useful proxy for coil complexity. For example, the ROSE code \citep{drevlak} enables the inclusion of properties of the current potential in the objective function. In this work, we adopt a similar approach but enable the efficient computation of shape derivatives of such metrics for gradient-based optimization.

% Since the shape gradient describes the behavior of a functional on any given region of a surface, the shape gradient can be used to quantify the engineering tolerances to which the coils shapes must be built. Such tolerances have historically been a driving factor in the cost of stellarator construction.

The goal of this work is to understand which plasma boundary shapes are consistent with simple coils using the shape gradient, a scalar functional defined on the plasma boundary that quantifies the local sensitivity of an objective to perturbation of the surface. Specifically, we will consider an objective which quantifies coil complexity using the current potential solution on a winding surface. While coil complexity may be reduced by simultaneously optimizing the winding surface shape, for simplicity, we constrain the winding surface to be uniformly offset from the plasma surface.

Other approaches have been used to understand the relationship between the complexity of external currents and magnetic surface shapes. A singular value decomposition of the matrix coupling external fields and the normal field on the plasma boundary can quantify the ``efficiency'' of an external magnetic field on a control surface in producing a given plasma boundary \citep{landreman2016}. This analysis highlights certain characteristics that are difficult to produce with external magnetic fields, such as boundaries with small-wavelength features and concavity. By considering a two-parameter family of magnetic surfaces near the magnetic axis with fixed rotational transform, it has also been observed that ellipticity of the boundary tends to increase the coil complexity more than the torsion of the axis \citep{hudson}. We present a different approach to put plasma surface optimization and coil design on a similar footing. By determining how the coil complexity metrics computed from REGCOIL depend on perturbations of the plasma boundary, plasma shapes can be designed which give rise to simpler coils. 

Using the derivatives of the objective function with respect to the shaping parameters, or the shape derivatives, we can construct the shape gradient by solving a linear system \citep{Landreman2018}. 
% The shape gradient is a scalar function defined on the plasma boundary which quantifies the local sensitivity of a function with respect to the normal perturbation of the shape. 
The shape gradient can be used for efficient gradient-based optimization of the shape. Furthermore, the evaluation of the shape gradient enables the identification of features of the plasma boundary that are amenable to simpler coils. Using techniques similar to those presented in this work, the shape gradient of coil complexity metrics with respect to perturbations of $S_{\text{coil}}$ has previously been computed \citep{Paul2018}. This analysis highlighted features of the winding surface which allow for more accurate reproduction of the desired plasma surface, and gradient-based optimization of the winding surface was demonstrated.

Stellarators typically require a large set of parameters to fully describe their three-dimensional geometry. The optimization in REGCOIL introduces an implicit dependence of the metrics on the plasma surface parameters through the linear least-squares system. This means that taking these derivatives with a finite-difference method would ordinarily require solving a linear system similar to the one solved by REGCOIL corresponding to the perturbation of each surface parameter; however, this can be avoided using the adjoint method. A variable, called the adjoint variable, can be computed by solving a similar linear system. However, the adjoint variable is common to all the surface parameters, so the system only needs to be solved one additional time for all the parameters. With the application of an adjoint method, we furthermore eliminate the noise associated with the finite-difference step size. The adjoint method has recently been applied to several problems in stellarator design \citep{Paul2018,Paul2019,Paul2020,Antonsen2019,Giuliani2020,Geraldini2021} and is commonly implemented in the fields of fluid dynamics and aerodynamic engineering \citep{Jameson1998,Othmer2008}.

In \S\ref{sec_regcoil}, we will give an overview of the current potential method used in the REGCOIL code. We provide some background on the theory of shape gradients and their application to stellarator optimization in \S\ref{sec_shape_grad}. We compute the shape gradient with a specific choice of poloidal angle described in \S\ref{sec:poloidal_angles} which is chosen for its spectral condensation. To construct the shape gradients, the derivatives of the REGCOIL metrics with respect to the plasma surface parameters are described in \S\ref{sec_plasma_derivs}. In \S\ref{sec_benchmarks}, benchmarks of these derivatives and shape gradients computed from them are shown. Finally, the shape gradients of the figures of interest are also shown in \S\ref{sec_benchmarks} for the W7-X \citep{Beidler1990}  and QHS46 \citep{qhs46} configurations.

\section{Overview of the current potential method}
\label{sec_regcoil}
 When optimizing coils, a balance must be reached between a simple coil design and accuracy in reproducing the desired plasma shape. REGCOIL \citep{landreman2017} uses a current potential approximation to formulate this optimization as a linear least-squares system with a regularization factor balancing the coil complexity and plasma surface accuracy. The surface current on the coil winding surface is taken to be of the form,
\begin{equation}
    \boldsymbol{K}=\boldsymbol{\hat{n}}' \times \bnabla \Phi,
    \label{surf_cur}
\end{equation}
where $\Phi$ is the current potential and $\boldsymbol{\hat{n}}'$ is the outward unit normal vector to the coil surface. The surface current is then divergence-free and tangential to the coil surface. Since the gradient and level curves of the current potential are perpendicular and on the coil surface, the surface current flows along the level curves of the current potential. The discrete coil shapes are then determined by taking a selection of level curves of the current potential, or streamlines of $\bm{K}$. These level curves will be closer together in regions of high current density, implying a reduction of the discrete coil-coil separation. 

This current potential has secular terms
% , which increase continually as the angles which map the coil surface increase, 
and a single-valued term. The single-valued component of the current potential can be expanded as a sine series over the coil surface under the assumption of stellarator symmetry \citep{Dewar1998}. The current potential then has the form,
\begin{equation}
    \Phi = G\frac{\zeta'}{2\upi} + I\frac{\theta'}{2\upi} + \sum_j \Phi_j \sin(m_j\theta' - n_j\zeta'),
    \label{cur_pot}
\end{equation}
where $G$ and $I$ are currents which link the coil surface poloidally and toroidally, $\theta'$ is a poloidal angle, and $\zeta'$ is a toroidal angle. 

We consider two figures of merit,
\begin{align}
\begin{split}
    \chi^2_B &=\int_{S_{\text{plasma}}} \left(\boldsymbol{B}(\theta,\zeta)\bcdot \boldsymbol{\hat{n}}(\theta,\zeta) \right)^2 \mathrm{d}a \label{eq:chi2_B} \\
    \chi^2_K &=\int_{S_{\text{coil}}} \left|\boldsymbol{K}(\theta',\zeta') \right|^2 \mathrm{d}a',
\end{split}
\end{align}
where $\boldsymbol{B}$ is the magnetic field and $\boldsymbol{\hat{n}}$ is the outward unit normal vector to the plasma surface. 
% The plasma and coil surfaces are parameterized by two angles, a poloidal angle $\theta$ which goes the short way around the torus and a toroidal angle $\zeta$ which goes the long way around the torus. 
In keeping with \cite{landreman2017}, primed coordinates refer to the coil surface while unprimed coordinates refer to the plasma surface. Similarly, primed and unprimed quantities are used to refer to those of the coil and plasma surfaces, respectively, when a distinction is necessary. The integral for $\chi^2_B$ is over the desired magnetic surface, $S_{\text{plasma}}$, and the integral for $\chi^2_K$ is over the coil winding surface, $S_{\text{coil}}$. The goal is for $S_{\text{plasma}}$ to be a magnetic surface, so the normal component of the magnetic field will vanish if the plasma surface is properly reconstructed and, thus, so will $\chi^2_B$.  Large values of $\chi^2_K$ are correlated with increased coil complexity \citep{Paul2018}. 
% The more the coils depart from planar curves, the larger $\chi^2_K$ will be. 
% In addition to this, for the surface current defined by (\ref{surf_cur}), %and the choice of discrete wires from contours of the continuous current potential, 
Regions of higher surface current density require reduced coil-coil spacing; thus, the coil-coil separation is increased by reducing $\chi^2_K$.

We make the approximation that the normal field from the plasma current is negligible such that only the vacuum field from the coils is included in \eqref{eq:chi2_B}. In \S 6, examples will be shown for the vacuum QHS46 and W7-X equilibria. If the derivatives of the normal field from the plasma current with respect to the plasma boundary are computed \citep{Henneberg2020}, the adjoint method we present in \S \ref{sec_plasma_derivs} could be modified to include the dependence of $\chi^2_B$ on the plasma boundary through the plasma normal field.

% Large coil curvature and small coil-coil spacing makes construction and maintenance more difficult. 
When comparing configurations, it is also useful to consider normalized quantities,
% calculated from $\chi^2_B$ and $\chi^2_K$. These are given by,
\begin{equation}
\begin{split}
    \|B_n\|_2 &= \sqrt{\frac{\chi^2_B}{B_0^2 a_{\text{plasma}}}} \\
    \|K\|_2 &= \sqrt{\frac{\chi^2_K}{a_{\text{coil}}}},
\end{split}
\label{eq:norm_BK}
\end{equation}
where $a_{\text{plasma}}$ and $a_{\text{coil}}$ are the plasma and coil surface areas, respectively, and $B_0$ is a constant with the dimensions of magnetic field. In this work, we take it to be the root-mean-squared average over the plasma surface of the magnetic field strength from the equilibrium. 

REGCOIL minimizes the quantity $\chi^2_B + \lambda \chi^2_K$ by solving a linear least-squares system for the Fourier modes of the single-valued current potential, $\Phi_j$, defined in (\ref{cur_pot}). Here $\lambda$ is the regularization parameter which sets the relative importance of $\chi^2_B$ and $\chi^2_K$ in the optimization. A fixed value of $\lambda$ can be chosen, or the freedom in $\lambda$ can be use to fix a certain target function, such as $\chi^2_K$ or $\|K\|_2$. We employ the notation  $\boldsymbol{\Phi}$ to represent the vector of unknowns, $\{\Phi_j\}$. The linear least-squares system takes the form,
\begin{equation}
    \mathsfbi{A}\boldsymbol{\Phi} = \boldsymbol{b},
    \label{sol_sys}
\end{equation}
with,
\begin{equation}
\begin{split}
    \mathsfbi{A}&=\mathsfbi{A}^B+\lambda \mathsfbi{A}^K\\
    \boldsymbol{b}&=\boldsymbol{b}^B+\lambda \boldsymbol{b}^K,
\end{split}
\label{eq:A_b}
\end{equation}
where the $B$ matrix and vector are associated with the system that optimizes only $\chi^2_B$, and the $K$ matrix and vector are associated with the system that optimizes only $\chi^2_K$. 
% Both of these matrices are square. 
Expressions for these quantities are given in Appendix A of \citep{landreman2017}.

\section{Shape gradients}
\label{sec_shape_grad}

% The traditional approach to stellarator design is to first choose a plasma shape for its magneto-hydrodynamic (MHD) properties then to design the coil shapes to reproduce the desired plasma boundary. This paper presents a different approach to put plasma and coil surface design on a similar footing. By determining how the coil complexity metrics of REGCOIL depend on perturbations of the plasma boundary, plasma shapes can be designed which give rise to simpler coils. 

The shape gradient describes the local dependence of a functional on a surface and is independent of the choice of basis or parameterization. Consider a functional $F$ of the plasma surface, such as $\chi^2_B$, $\chi^2_K$,  $\|B_n\|_2$, or $\|K\|_2$. The Hadamard-Zol\'{e}sio structure theorem \citep{hadamard} states that the change in $F$ due to a vector field of infinitesimal displacements $\delta \boldsymbol{r}$ to the plasma surface $S_{\text{plasma}}$ is given by,
\begin{equation}
    \delta F(S_{\text{plasma}};\delta \boldsymbol{r}) = \int_{S_{\text{plasma}}} S_F \delta \boldsymbol{r} \bcdot \boldsymbol{\hat{n}} \ \mathrm{d}a,
    \label{eq:shape_gradient}
\end{equation}
under the assumption that $\delta F$ exists for all infinitely differentiable and compactly supported $\delta \bm{r}$. Here $S_F$ is the shape gradient, and $\delta F$ is called the shape derivative. In this equation and the following ones, the integral is over the unperturbed plasma surface and $\boldsymbol{\hat{n}}$ is the unit normal outward to this surface. This gives information about how the plasma surface can be deformed to alter a functional of the surface. For the case that the functional $F$ is $\chi^2_K$ or $\|K\|_2$, its shape gradient enables the plasma surface to be deformed to obtain simpler coils. 
% This is of great importance to the practicality of constructing a stellarator.
% The shape gradient can also be used to set engineering tolerances on coil shapes since it relates changes in a functional of interest to changes in small regions of the boundary. 
% Coil tolerances are an important factor in stellarator design which drives the cost of construction. 

Given a parameterization of the plasma surface with a set of parameters $\boldsymbol{\Omega}$, \eqref{eq:shape_gradient} can be expressed as,
% depends on the parameterization chosen, of
% this equation is,
\begin{equation}
    \frac{\p F}{\p \Omega_j} = \int_{S_{\text{plasma}}} S_F \frac{\p \boldsymbol{r}}{\p \Omega_j} \bcdot \boldsymbol{\hat{n}} \ \mathrm{d}a.
\end{equation}
The shape gradient can be expanded as a Fourier series, assuming stellarator symmetry, as,
\begin{equation}
    S_F=\sum_i S_F^i \cos(m_i \theta - n_i \zeta),
\end{equation}
% The condition for stellarator symmetry is given in Eq.~. 
% The derivative of $F$ with respect to one of the plasma surface parameters can then be written as,
such that the linear system for the shape gradient Fourier coefficients, $\{S_F^i\}$, is,
\begin{equation}
    \frac{\p F}{\p \Omega_j} = \sum_i S_F^i \int_{S_{\text{plasma}}} \cos(m_i \theta - n_i \zeta) \frac{\p \boldsymbol{r}}{\p \Omega_j} \bcdot \boldsymbol{\hat{n}} \ \mathrm{d}a.
    \label{shape_grad_lin_sys}
\end{equation}
% This equation is now a linear system in the shape gradient Fourier coefficients $S_F^i$. 
If the same number of modes are used to discretize $S_F$ as the number of modes retained in $\{\partial F/\partial \Omega_j\}$, then the system is square, and the shape gradient can then be constructed from a local parameterization of the surface with a linear solve. If the system is not square, it can still be solved \citep{Landreman2018} using a pseudoinverse or QR factorization. Details of the chosen parameterization of the plasma boundary are provided in the following Section.

% In this paper, a set of Fourier amplitudes are used to parameterize the plasma surface, which are defined in (\ref{l_terms}). Two possible parameterizations are considered here. The first is the Fourier expansion of the cylindrical coordinates, $R$ and $Z$. The second is the Fourier expansion of the distance from a fixed axis using a poloidal angle defined by the arctangent of $R$ and $Z$. To construct a shape gradient of one of the REGCOIL metrics, derivatives of that metric must be taken with respect to the Fourier coefficients used to describe the plasma surface. 

\section{Efficient Fourier representation of the plasma boundary}
\label{sec:poloidal_angles}

A toroidal surface such as $S_{\text{plasma}}$ can be specified by two Fourier series in the cylindrical coordinates $R$ and $Z$,
\begin{equation}
\begin{split}
    R(\theta,\zeta)&=\sum_{m,n}R_{m,n}^c \cos(m\theta-n\zeta)\\
    Z(\theta,\zeta)&=\sum_{m,n}Z_{m,n}^s \sin(m\theta-n\zeta).
\end{split}
\label{eq:double_Fourier}
\end{equation}
Here $\zeta$ is taken to be the cylindrical toroidal angle and $\theta$ is a poloidal angle. As a Fourier series in two separate variables is required, we will refer to this as the double Fourier representation. This representation is used in the VMEC
% used by VMEC, which is chosen for efficient convergence of the Fourier series 
\citep{hirshman} and SPEC \citep{Hudson2012} equilibrium codes. Here we have assumed stellarator symmetry such that,
\begin{equation}
    \begin{split}
        &R(-\theta,-\zeta)=R(\theta,\zeta)\\
        &Z(-\theta,-\zeta)=-Z(\theta,\zeta).
    \end{split}
    \label{stell_sym}
\end{equation} 
Because this representation does not constrain the poloidal angle, the Fourier amplitudes could be altered corresponding to a redefinition of the poloidal angle, while the surface is left unchanged. This freedom in the choice of poloidal angle has been used to condense the Fourier spectrum using a variational method \citep{Hirshman1985,Hirshman1998}. In the context of fixed-boundary optimization, a uniquely defined poloidal angle may eliminate a null space in the parameter space. 

To eliminate such a null space, we consider a representation which requires only one series representation of a radial distance, $l$, from a fixed coordinate axis \citep{Hirshman1998}. In this work, we take the coordinate axis to coincide with the magnetic axis from the equilibrium, although this assumption is not necessary. We define a unique poloidal angle $\vartheta$ using the arctangent in the poloidal plane,
\begin{equation}
\begin{split}
    l=\sqrt{(R-R_0)^2+(Z-Z_0)^2} \\
    \vartheta=\arctan\left(\frac{Z-Z_0}{R-R_0}\right),
\end{split}
\end{equation}
where $R_0(\zeta)$ and $Z_0(\zeta)$ are the cylindrical coordinates of the axis. Note that a representation of this form assumes that each cross-section in the poloidal plane is a star domain, indicating that there exists a coordinate axis $(R_0,Z_0)$ such that the line segment connecting the axis and any point on the boundary is contained within the boundary.

\begin{figure}	%---------------------------------
\centering 
\begin{subfigure}[b]{0.49\textwidth}
\centering
\includegraphics[trim=14cm 1cm 16cm 3cm, clip, width=1.0\textwidth]{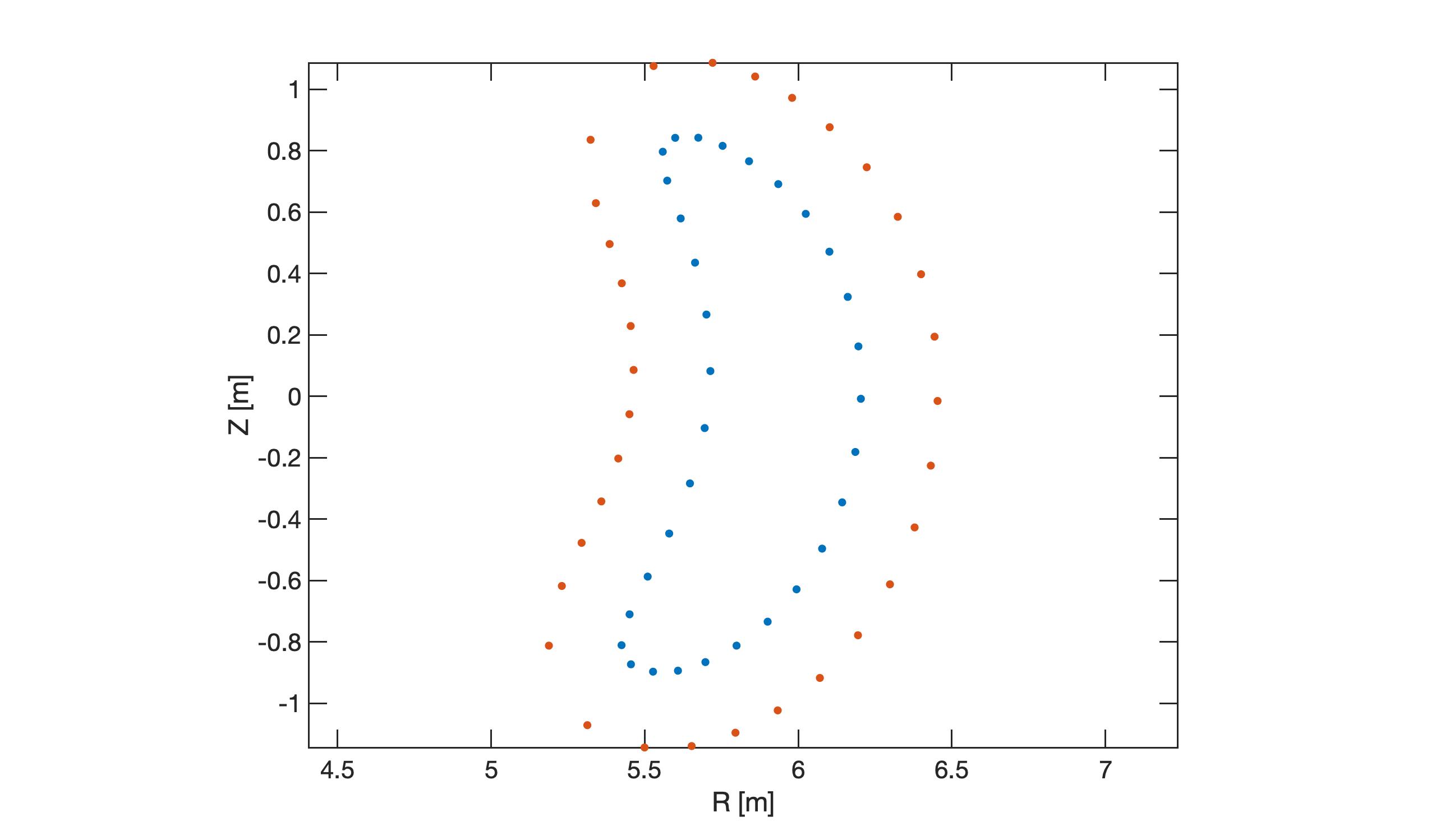}
\caption{}
\label{cross_section_vmec}
\end{subfigure}
\begin{subfigure}[b]{0.49\textwidth}
\centering
\includegraphics[trim=14cm 1cm 16cm 3cm, clip, width=1.0\textwidth]{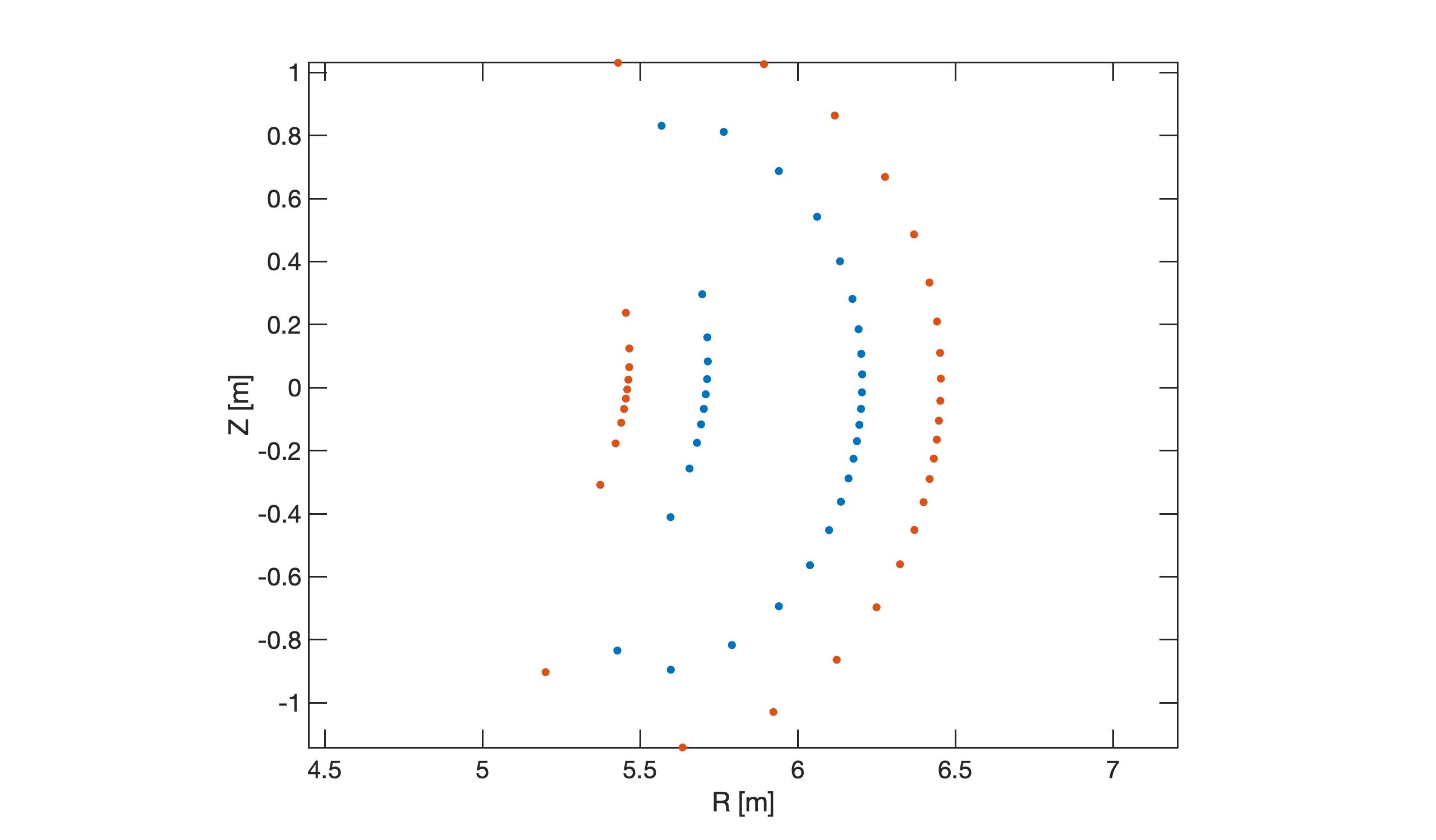}
\caption{}
\label{cross_section_vartheta}
\end{subfigure}
% next line
\begin{subfigure}[b]{0.49\textwidth}
\centering
\includegraphics[trim=14cm 1cm 16cm 3cm, clip, width=1.0\textwidth]{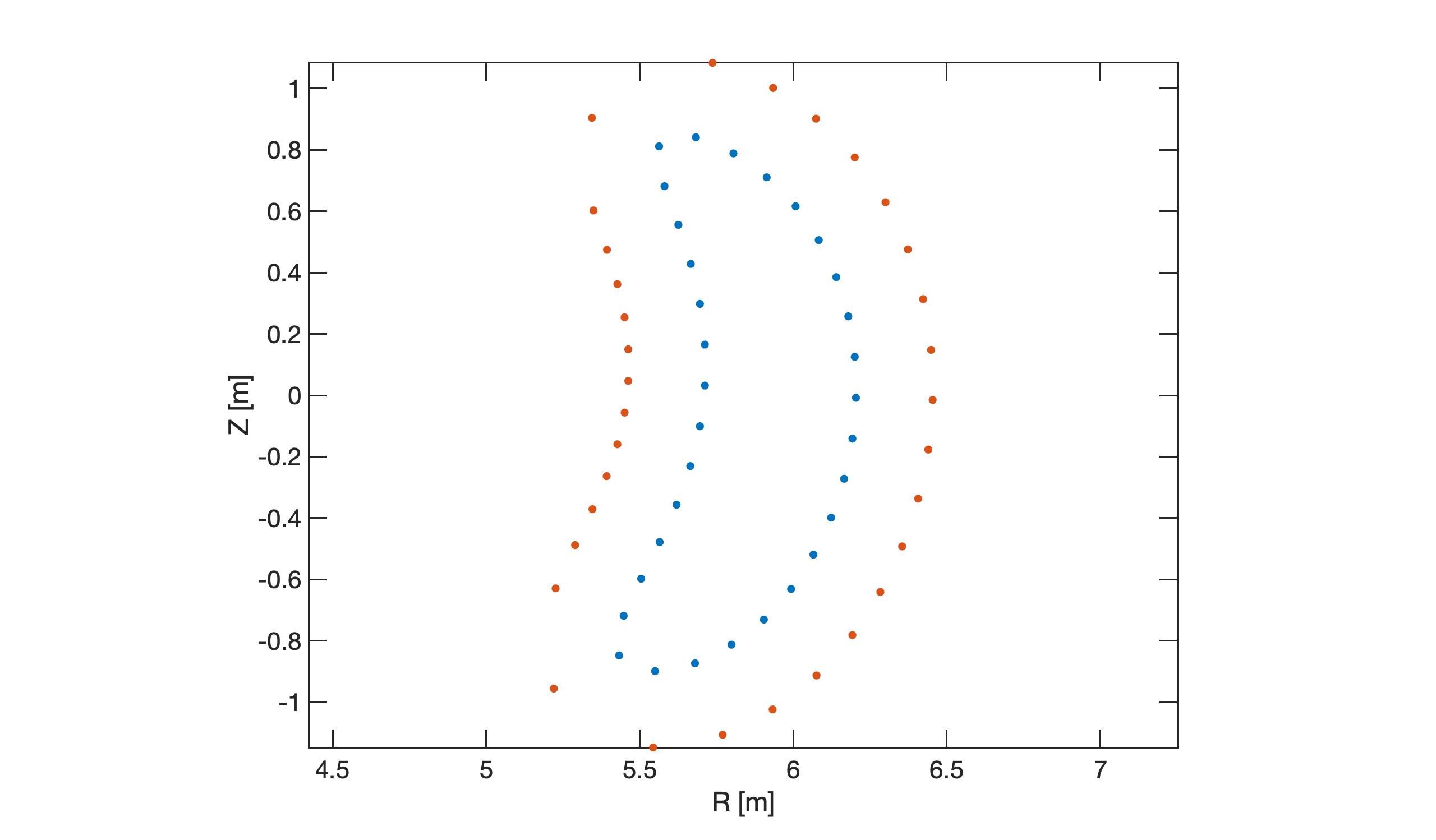}
\caption{}
\label{cross_section_arclength_p}
\end{subfigure}
\begin{subfigure}[b]{0.49\textwidth}
\centering
\includegraphics[trim=14cm 1cm 16cm 3cm, clip, width=1.0\textwidth]{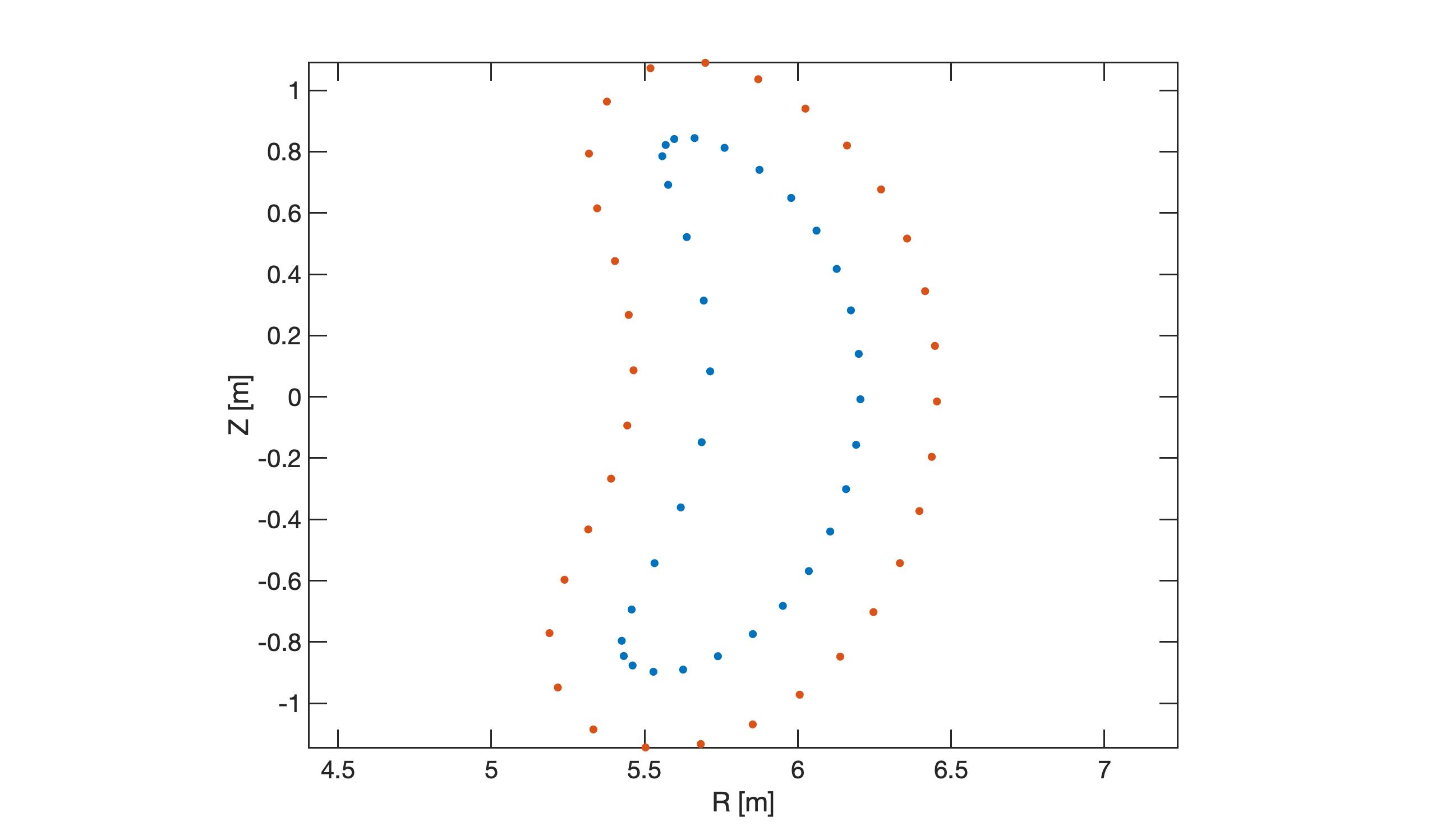}
\caption{}
\label{cross_section_arclength_c}
\end{subfigure}
% total caption
\caption{The bean-shaped cross-section of the W7-X configuration with a 0.25 m uniformly offset surface sampled at 32 values of the poloidal using (a) the initial VMEC representation, (b) the single Fourier representation with an arctangent angle, (c) the single Fourier representation with an arclength angle on the plasma surface, and (d) the single Fourier representation with an arclength angle on a surface uniformly offset from the plasma surface by 0.25 m.}
\label{cross_sections}
\end{figure}

The variable $l$ then admits a Fourier representation as,
\begin{equation}
    l(\vartheta,\zeta)=\sum_{m,n}l_{m,n}^c \cos(m\vartheta-n\zeta).
\label{l_series}
\end{equation}
% with,
% \begin{equation}
%     l_{m,n}^c=\frac{\int l(\vartheta,\phi)\cos(m\vartheta-n\zeta)\mathrm{d}\vartheta \mathrm{d}\zeta}{\int \cos^2(m\vartheta-n\zeta)\mathrm{d}\vartheta \mathrm{d}\zeta}.
%     \label{l_terms}
% \end{equation}
This representation uses a Fourier expansion in only a single quantity $l$ rather than in both $R$ and $Z$; for this reason, we will refer to this as the single Fourier representation.
% For the case of stellarator symmetry, only the cosine terms are needed. 
The cylindrical coordinates can now be expressed in the new representation as,
\begin{equation}
\begin{split}
    &R(\vartheta,\zeta)=R_0(\zeta)+l(\vartheta,\zeta)\cos(\vartheta)\\
    &Z(\vartheta,\zeta)=Z_0(\zeta)+l(\vartheta,\zeta)\sin(\vartheta).
\end{split}
\end{equation}
In practice, the Fourier series in \eqref{l_series} may decay rather slowly for relevant stellarator boundaries with widely varying curvature, as equally-spaced grid points in $\vartheta$ accumulate in regions of the surface where $l$ varies slowly with respect to $\vartheta$ and become sparse where $l$ varies rapidly with respect to $\vartheta$. In Figure \ref{cross_section_vartheta}, this accumulation occurs on the inboard and outboard sides of the plasma surface. 

% accumulate in concave regions of the surface and become sparse near the convex regions (Figure \ref{cross_sections}).

% In the rest of this paper, $\theta$ has been used rather than $\vartheta$ for simplicity. 

To improve the efficiency of this representation, a second unique poloidal angle can be defined by the arclength along the plasma surface at constant toroidal angle,
\begin{equation}
    \Bar{\theta}(\vartheta,\zeta)=2\pi \frac{\int_0^{\vartheta} \sqrt{\left(\frac{\p R}{\p \vartheta'}\right)^2 + \left(\frac{\p Z}{\p \vartheta'}\right)^2} \mathrm{d}\vartheta' }{\int_0^{2\pi} \sqrt{\left(\frac{\p R}{\p \vartheta'}\right)^2 + \left(\frac{\p Z}{\p \vartheta'}\right)^2} \mathrm{d}\vartheta' }.
\end{equation}
This choice of poloidal angle has been shown to correspond to the minimum of a spectral width functional \citep{Hirshman1998}. The Hirshman-Breslau technique can be used to obtain improved spectral convergence over the arclength angle by minimizing an energy functional with a nonlinear conjugate gradient method. In contrast, we present a method for improving the spectral convergence by computing an arclength angle on a uniform offset surface. This simply requires a Fourier transform to compute the coefficients in the new representation rather than the solution of a nonlinear optimization problem. We define a uniform offset surface by,
% The poloidal angle can be further modified to efficiently parameterize the plasma surface by computing an arclength angle on a uniform offset surface, defined by,
\begin{equation}
    \boldsymbol{r}_{\text{arclength}}=\boldsymbol{r}+a_{\text{arclength}}\boldsymbol{\hat{n}},
    \label{eq:angle_offset}
\end{equation}
where $a_{\text{arclength}}$ is a constant offset distance, taken normally to the plasma surface $\bm{r}$. This parameterization is particularly important for constraining the coil winding surface as described in \S\ref{sec:offset}. It is critical that the offset distance $a_{\text{arclength}}$ is not too large; otherwise, the offset surface will self-intersect. The maximum offset distance is constrained by the principal curvatures, $\kappa_1$ and $\kappa_2$, of $S_{\text{plasma}}$. Here we use the convention that $\kappa_{1,2}<0$ indicates concavity. The maximum curvature in the outwardly concave regions of the plasma surface ($\kappa_{1,2}<0$) determines the minimum outward radius of curvature, and thus the maximum outward ($a_{\text{arclength}}>0$) offset distance. Conversely, the maximum curvature in the outwardly convex regions of the plasma surface ($\kappa_{1,2}>0$) determines the maximum inward ($a_{\text{arclength}}<0$) offset distance \citep{Farouki1986}. The constraint on the offset distance is, therefore, given by,
\begin{align}
 -\frac{1}{\max\left\{\kappa_1, \kappa_2\right\}} <  a_{\text{arclength}} < \frac{1}{|\min\left\{\kappa_1, \kappa_2\right\}|}.
    % |a| < \min\left\{\frac{-\text{sign}(a)}{\kappa_1}, \frac{-\text{sign}(a)}{\kappa_2}\right\} .
    \label{eq:angle_max_offset}
\end{align}
Note that for any closed toroidal surface, the surface integral of the Gaussian curvature ($\int  \kappa_1 \kappa_2 \, d^2 a$) must vanish, indicating that the minimum of the principal curvatures must always be negative and the maximum of the principal curvatures must be positive. 

In this case, rather than using the arclength on the plasma surface to define $\Bar{\theta}$, the arclength is computed on a surface uniformly offset from the plasma boundary as defined in (\ref{eq:angle_offset}) with the restriction on outward offsets as in (\ref{eq:angle_max_offset}). In this way, the poloidal angle can be tuned to accommodate a particular offset of the coil surface from the plasma surface or to increase resolution in regions of large (positive or negative) curvature. 
% It is not necessary, though, to use the same offset to compute the arclength angle and the uniformly offset coil surface.

We define $\omega$ to be the difference between the arclength and arctangent angles,
\begin{equation}
    \omega(\vartheta,\zeta)=\Bar{\theta}(\vartheta,\zeta)-\vartheta.
\end{equation}
The quantities $l$ and $\omega$ can now be expressed as Fourier series in $\Bar{\theta}$,
\begin{equation}
    \begin{split}
        l(\Bar{\theta},\zeta)&=\sum_{m,n}l_{m,n}^c \cos(m\Bar{\theta}-n\zeta)\\
        \omega(\Bar{\theta},\zeta)&=\sum_{m,n}\omega_{m,n}^s \sin(m\Bar{\theta}-n\zeta).
    \end{split}
\label{l_series_arclength}
\end{equation}
% with,
% \begin{equation}
% \begin{split}
%     l_{m,n}^c&=\frac{\int l(\Bar{\theta},\phi)\cos(m\Bar{\theta}-n\zeta)\mathrm{d}\Bar{\theta} \mathrm{d}\zeta}{\int \cos^2(m\Bar{\theta}-n\zeta)\mathrm{d}\Bar{\theta} \mathrm{d}\zeta}\\
%     \omega_{m,n}^s&=\frac{\int l(\Bar{\theta},\phi)\sin(m\Bar{\theta}-n\zeta)\mathrm{d}\Bar{\theta} \mathrm{d}\zeta}{\int \sin^2(m\Bar{\theta}-n\zeta)\mathrm{d}\Bar{\theta} \mathrm{d}\zeta}.
% \end{split}
% \label{l_terms_arclength}
% \end{equation}
Note that while we express $\omega$ in a Fourier series to describe the relationship between the arctangent angle $\vartheta$ and the arclength angle $\Bar{\theta}$, it is only necessary to differentiate with respect to $l_{m,n}^c$ when constructing shape gradients. In this way, the poloidal angle is fixed as the plasma boundary is perturbed, and the corresponding null space is eliminated. The cylindrical coordinates can now be expressed in the new representation as,
\begin{equation}
\begin{split}
    &R(\Bar{\theta},\zeta)=R_0(\zeta)+l(\Bar{\theta},\zeta)\cos(\Bar{\theta}-\omega(\Bar{\theta},\zeta))\\
    &Z(\Bar{\theta},\zeta)=Z_0(\zeta)+l(\Bar{\theta},\zeta)\sin(\Bar{\theta}-\omega(\Bar{\theta},\zeta)).
\end{split}
\end{equation}
The arclength angle is advantageous over the arctangent angle because it allows for a more efficient representation of the plasma surface with a uniform sampling of the poloidal angle. This is particularly useful for representing plasma boundaries with regions of large curvature. In Figure \ref{cross_sections}, we compare the distribution of poloidal grid points on the bean-shaped cross-section of the W7-X boundary, which features both concave and convex regions. The grid points in the plasma arclength angle are equally-spaced along the plasma surface cross-section while the points in the arctangent angle accumulate on the inboard and outboard sides. Observe that in highly convex regions of the plasma surface, the normal vector is changing rapidly, causing points on the offset surface to diverge from each other. By fixing the points on the offset surface to be spaced by uniform arclength increments, the points on the plasma surface tend to accumulate in the highly convex regions. Extra resolution in these regions is beneficial for improved convergence of REGCOIL calculations as well as a more efficient Fourier series representation. Alternatively, for surfaces which feature highly concave regions, it may be more advantageous to use an inward offset when computing the arclength angle. In this case, the maximum inward offset is set by the maximum curvature in the outwardly convex regions as in (\ref{eq:angle_max_offset}).

% Compare figure \ref{cross_section_vartheta} and figure \ref{cross_section_arclength_p} to see how the choice of poloidal angle affects the distribution of points on the plasma and uniformly offset coil surfaces for the single Fourier series representation.
	%---------------------------------

%  See figure \ref{cross_sections} for a comparison of the distribution of points on the plasma and uniformly offset coil surfaces for the different representations discussed. Using an offset surface to compute the arclength angle also directly benefits the plasma surface. 
\begin{figure}	%---------------------------------
\centering 
\begin{subfigure}[b]{0.44\textwidth}
\hspace*{0cm}\includegraphics[width=1.0\textwidth]{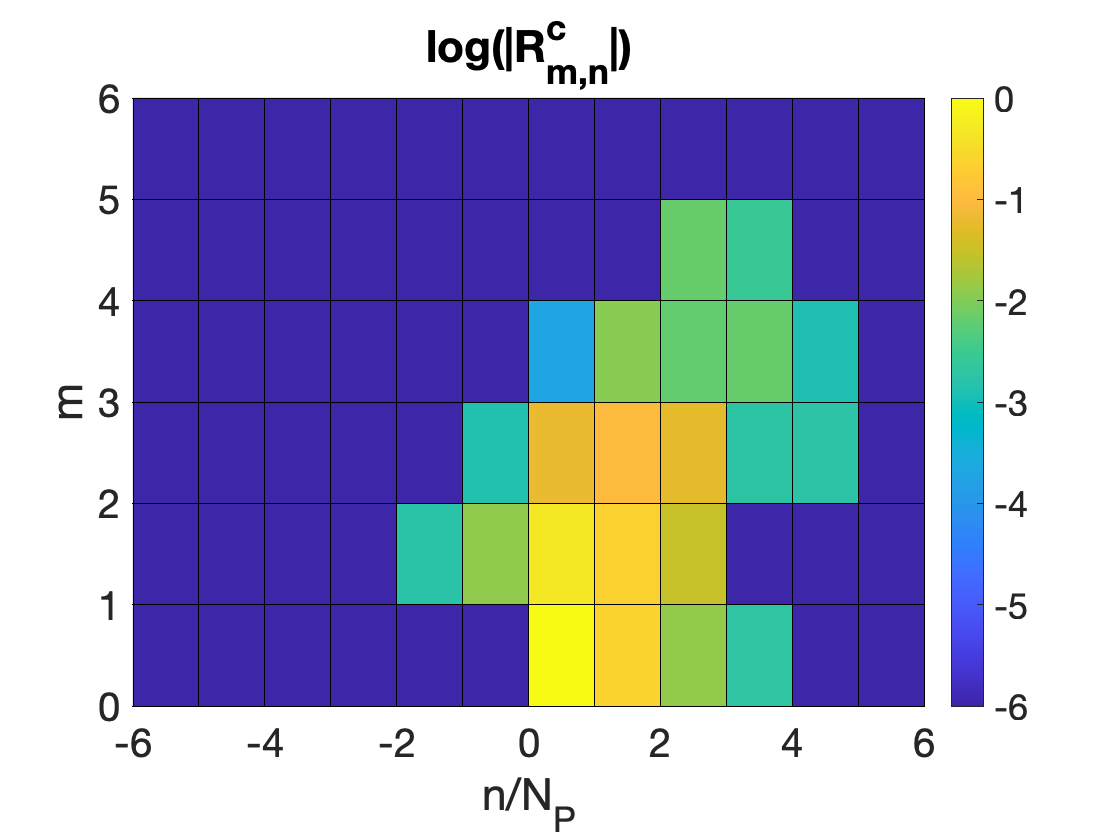}
\caption{}
\label{rmnc_pcolor}
\end{subfigure}
\begin{subfigure}[b]{0.44\textwidth}
\hspace*{0cm}\includegraphics[width=1.0\textwidth]{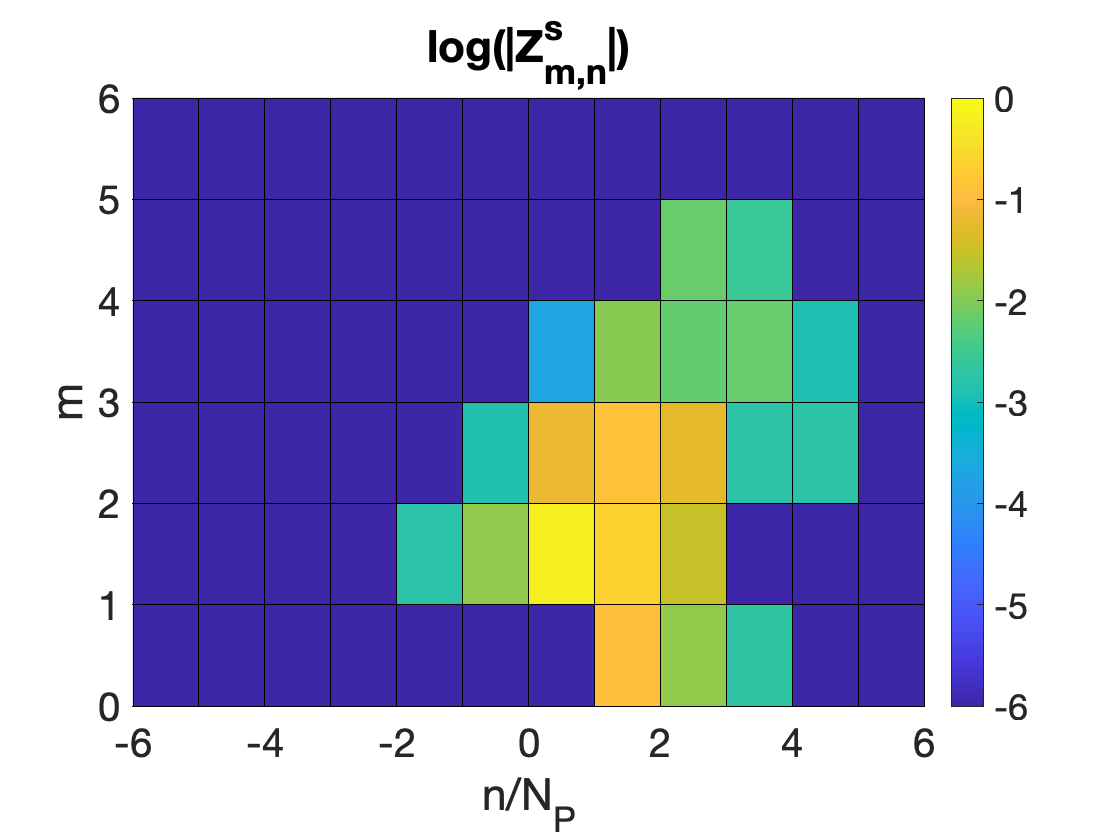}
\caption{}
\label{zmns_pcolor}
\end{subfigure}
\caption{ Magnitudes of (a) $R_{m,n}^c$ and (b) $Z_{m,n}^s$ for the W7-X boundary on a grid of mode numbers $m$ and $n/N_p$,  where $N_P$ is the number of field periods.}
\label{RZ_pcolor}
\end{figure}	%---------------------------------

\begin{figure}	%---------------------------------
\centering 
\begin{subfigure}[b]{1.0\textwidth}
\centering
\includegraphics[
% trim=3cm 13cm 4cm 14cm, clip,
width=0.4\textwidth]{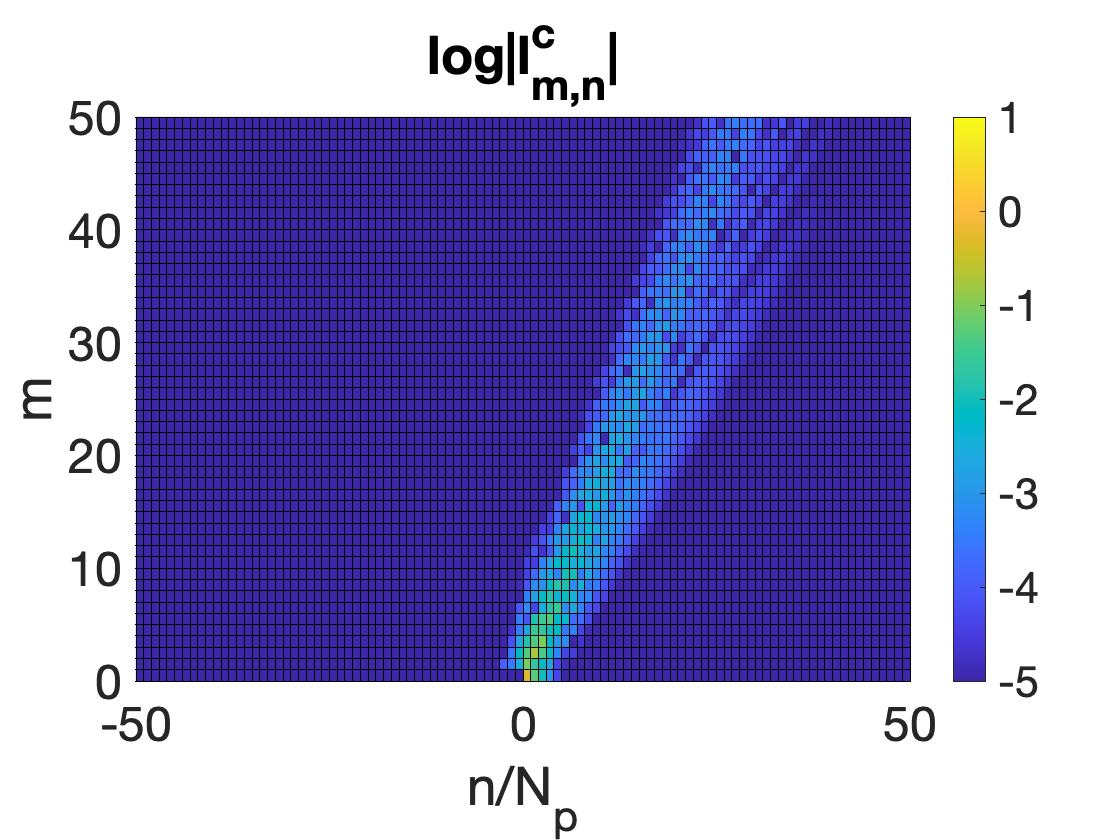}
%{lmnc_vartheta_spectrum}
\caption{}
\label{lmnc_pcolor}
\end{subfigure}
\begin{subfigure}[b]{0.4\textwidth}
\hspace*{0cm}\includegraphics[
% trim=3cm 13cm 4cm 14cm,
clip,width=1.0\textwidth]{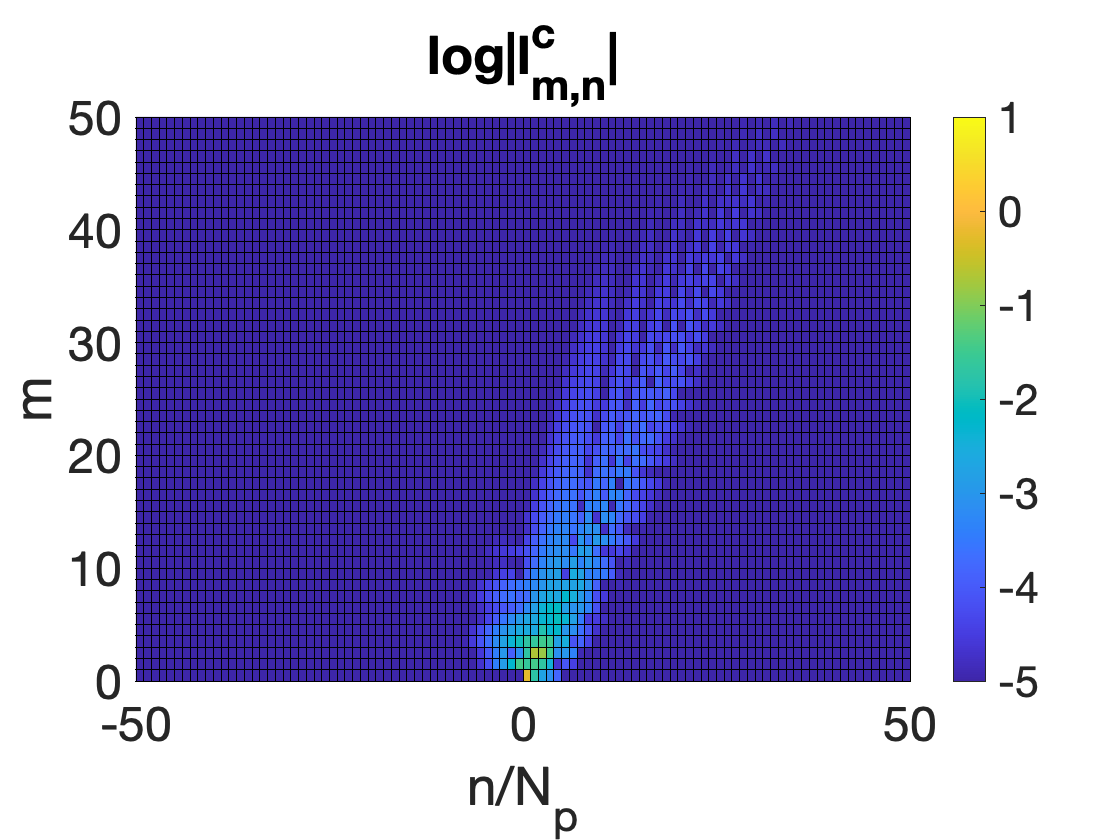}
% {lmnc_arclength_plasma_spectrum}
\caption{}
\label{lmnc_arclength_plasma_pcolor}
\end{subfigure}
\begin{subfigure}[b]{0.4\textwidth}
\hspace*{0cm}\includegraphics[
% trim=3cm 13cm 4cm 14cm, clip,
width=1.0\textwidth]{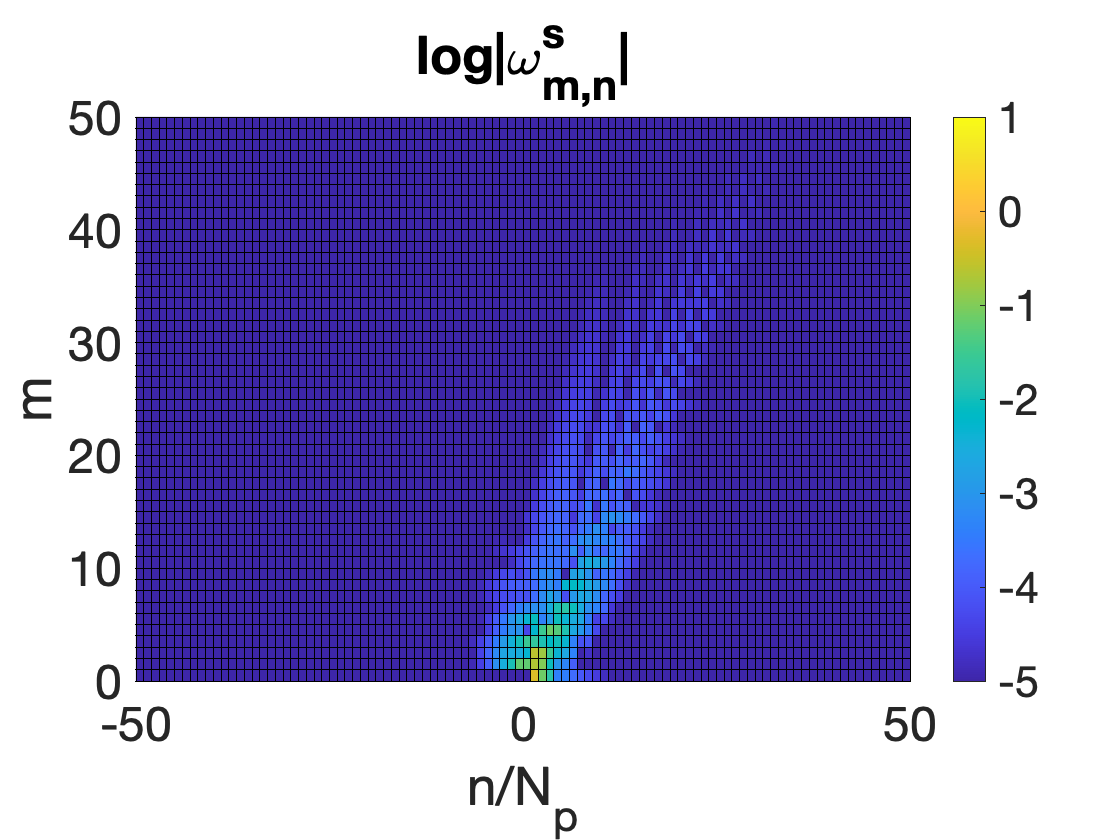}
% {omns_arclength_plasma_spectrum}
\caption{}
\label{omns_arclength_plasma_pcolor}
\end{subfigure}
\begin{subfigure}[b]{0.4\textwidth}
\hspace*{0cm}\includegraphics[
% trim=3cm 13cm 4cm 14cm, clip,
width=1.0\textwidth]{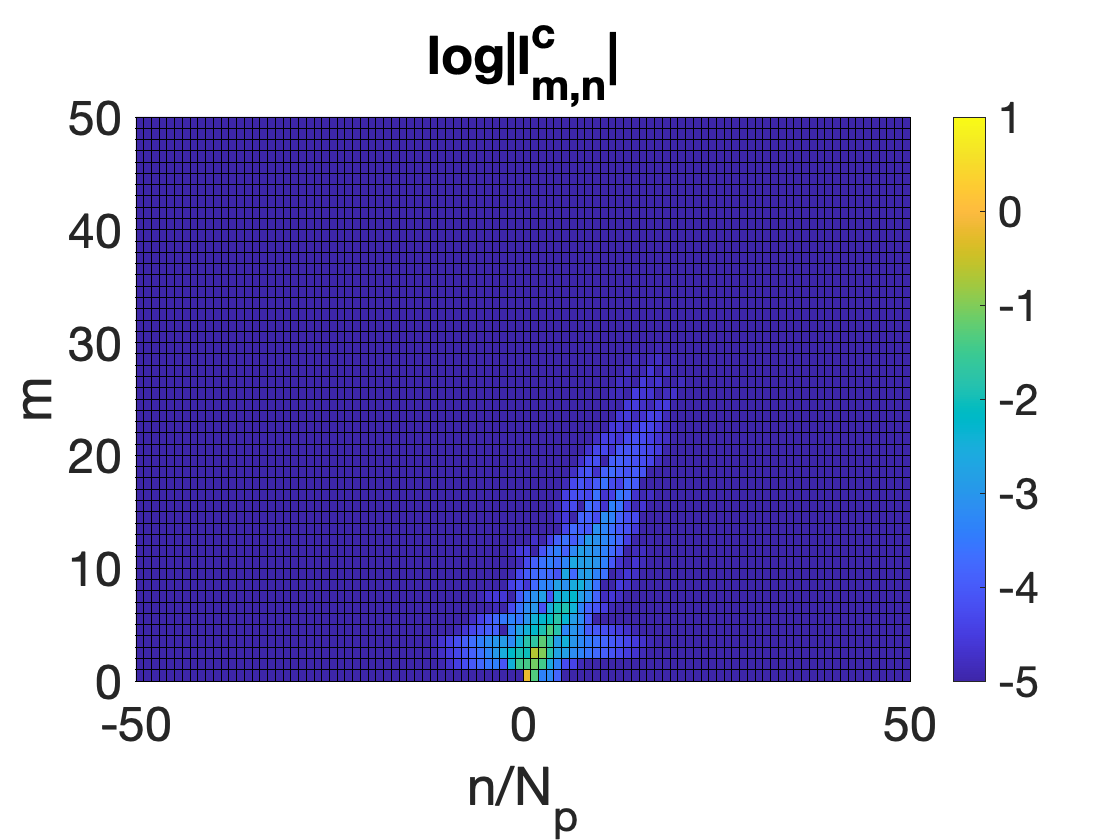}
% {lmnc_arclength_coil_spectrum}
\caption{}
\label{lmnc_arclength_coil_pcolor}
\end{subfigure}
\begin{subfigure}[b]{0.4\textwidth}
\hspace*{0cm}\includegraphics[
% trim=3cm 13cm 4cm 14cm, clip,
width=1.0\textwidth]{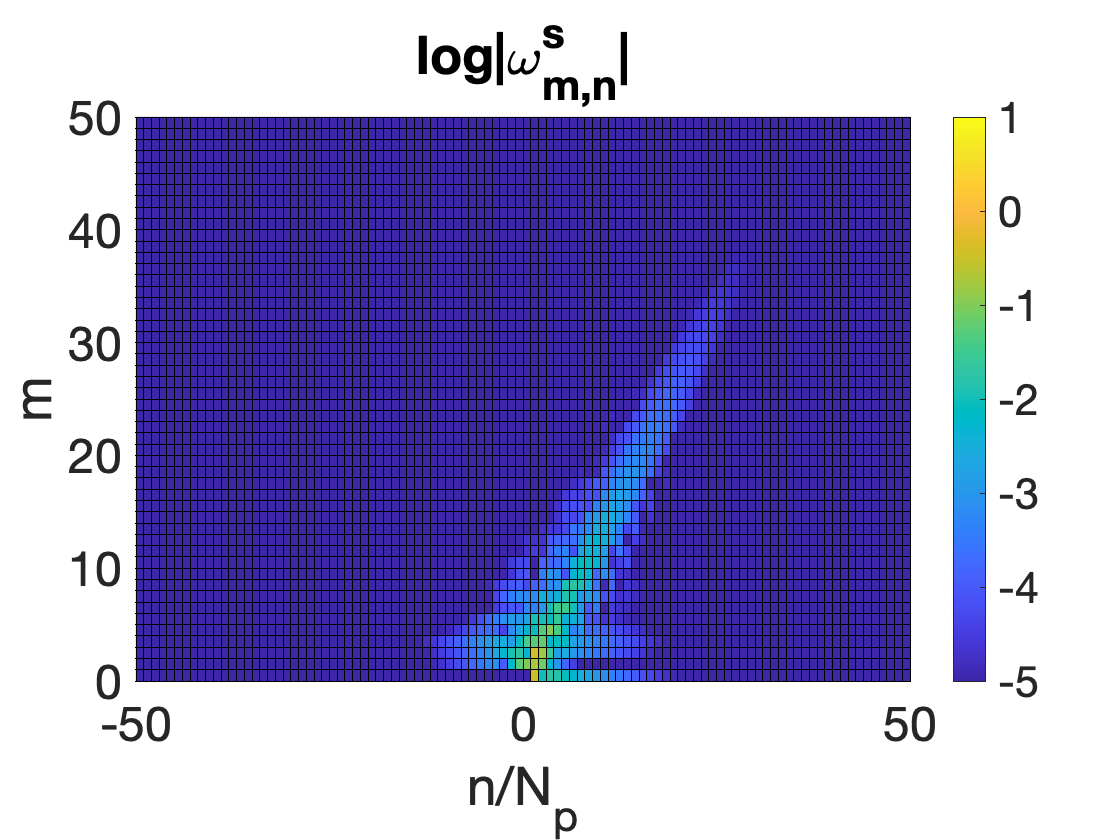}
\caption{}
\label{omns_arclength_coil_pcolor}
\end{subfigure}
\caption{(a) Magnitudes of $l_{m,n}^c$ for the arctangent poloidal angle on a grid of mode numbers $m$ and $n/N_P$. The dominant $l_{m,n}^c$ lie on a long, tilted band. Magnitudes of (b) $l_{m,n}^c$ and (c) $\omega_{m,n}^s$ for the arclength poloidal angle on the plasma surface. The amplitudes are more localized to the smaller mode numbers than with the arctangent angle, but still lie on a long, tilted band. Magnitudes of (d) $l_{m,n}^c$ and (e) $\omega_{m,n}^s$ for the arclength poloidal angle on a surface uniformly offset from the plasma surface by 0.25 m. The amplitudes are even more localized to the smaller mode numbers.}
\label{l_pcolor}
\end{figure}	%---------------------------------

% One limitation of this representation is that the modes do not decay for large $m$ and $n$ as they do for the double Fourier representation. 
To further evaluate these representations, we compare the convergence of the Fourier series for the W7-X surface. The double Fourier series representation \eqref{eq:double_Fourier} is compact in both $m$ and $n$  as shown in Figure \ref{RZ_pcolor}, as the surface was optimized with a truncated spectrum. (The surface is from a fixed-boundary equilibrium which predated the coil optimization.) For the other representations, the modes form a tilted band which is thin in the $n$ direction and decays slowly in the $m$ direction as shown in Figure \ref{l_pcolor}. With the arctangent representation, Figure \ref{lmnc_pcolor}, many modes must be retained in the Fourier series to accurately reconstruct the plasma surface. With the arclength angle defined on the plasma boundary, Figure \ref{lmnc_arclength_plasma_pcolor}, the magnitude of the Fourier modes decreases more rapidly in $m$. This decay is enhanced when the arclength angle is computed on a surface displaced from the plasma by 0.25 m, Figure \ref{lmnc_arclength_coil_pcolor}. A similar uniformly offset surface will be used in Section \ref{sec:shape_gradient_metrics} for computation of the shape gradients.
% which requires very high resolution in the $\theta$ and $\zeta$ grid, ; however, only those in the thin band are dominant. 
% This leads to high frequency noise in the reconstruction. This can be improved by filtering out modes smaller than a certain tolerance, which also gives the opportunity to reduce the data storage size.

\begin{figure}	%---------------------------------
\centering 
\begin{subfigure}[b]{0.4\textwidth}
\centering
\includegraphics[
% trim=3cm 13cm 4cm 14cm, clip,
width=1.0\textwidth]{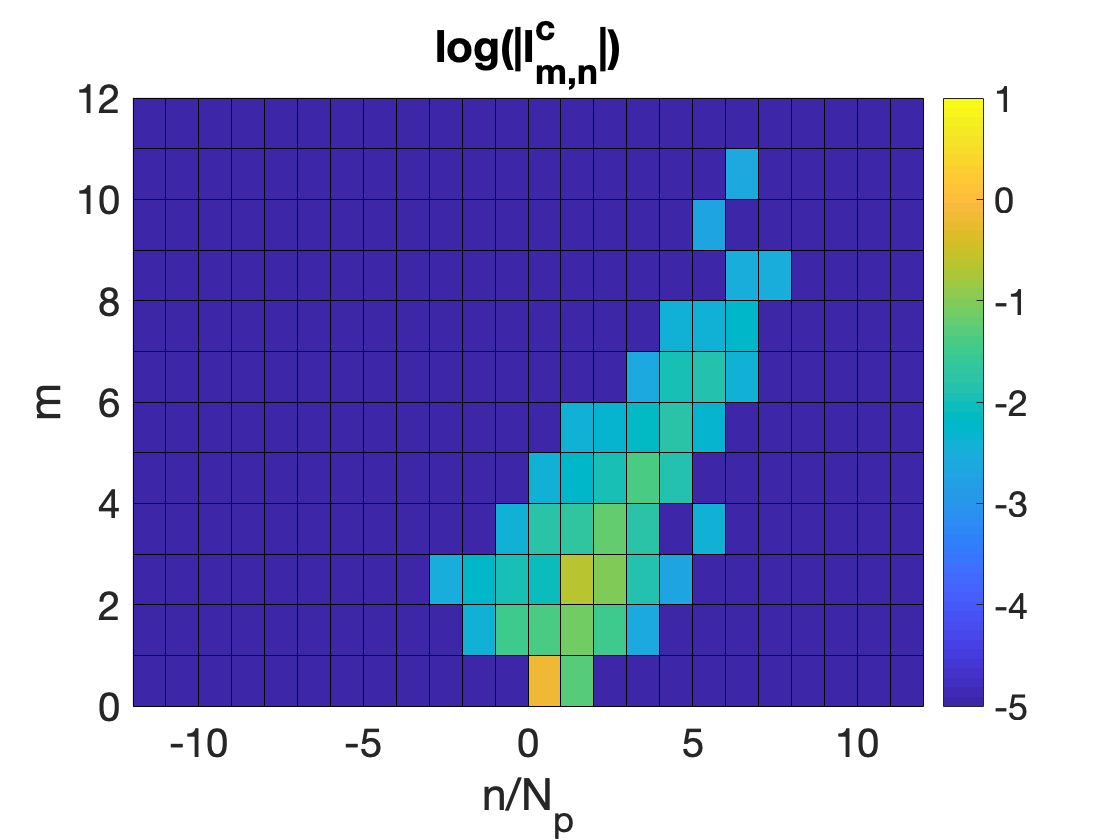}
%{lmnc_vartheta_spectrum}
\caption{}
\label{lmnc_condensed_pcolor}
\end{subfigure}
\begin{subfigure}[b]{0.4\textwidth}
\hspace*{0cm}\includegraphics[
% trim=3cm 13cm 4cm 14cm,
clip,width=1.0\textwidth]{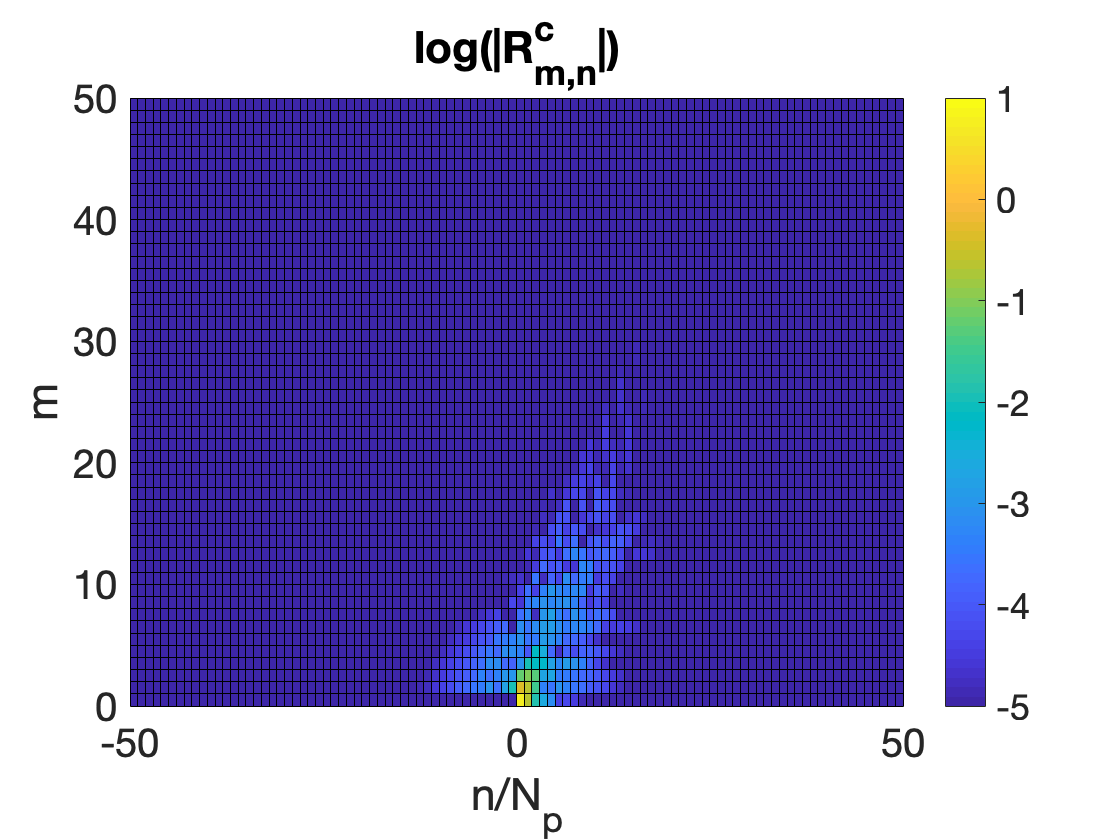}
% {lmnc_arclength_plasma_spectrum}
\caption{}
\label{rmnc_recon_pcolor}
\end{subfigure}
\begin{subfigure}[b]{0.4\textwidth}
\hspace*{0cm}\includegraphics[
% trim=3cm 13cm 4cm 14cm, clip,
width=1.0\textwidth]{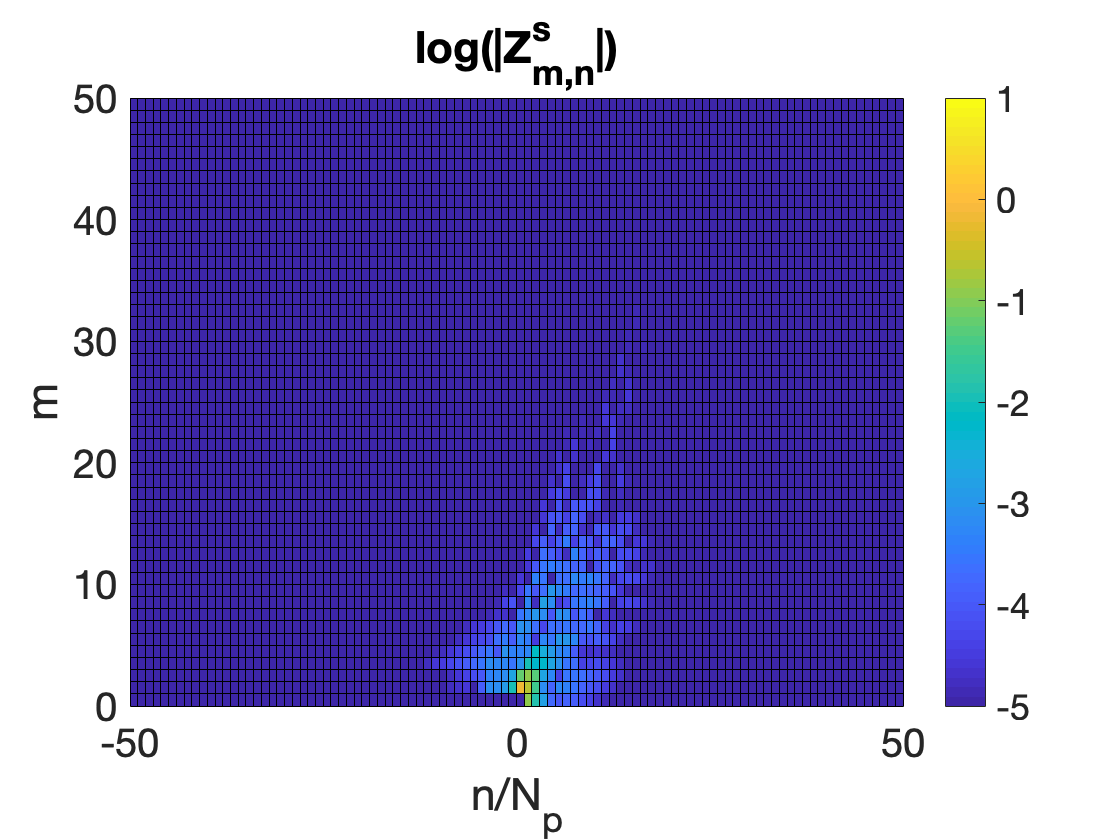}
% {omns_arclength_plasma_spectrum}
\caption{}
\label{zmns_recon_pcolor}
\end{subfigure}
\begin{subfigure}[b]{0.4\textwidth}
\centering
\includegraphics[
% trim=3cm 13cm 4cm 14cm, clip,
width=1.0\textwidth]{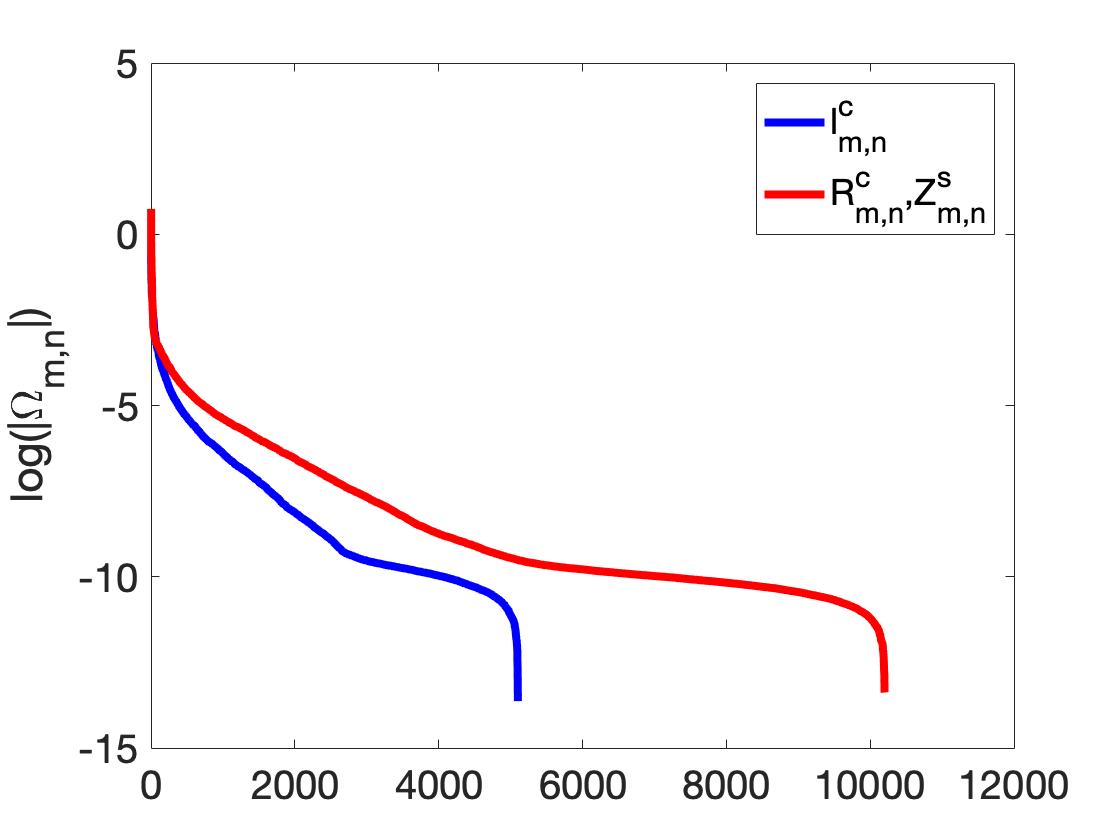}
%{lmnc_vartheta_spectrum}
\caption{}
\label{mode_decay}
\end{subfigure}
\caption{ Magnitudes of (a) $l_{m,n}^c$ for the arclength poloidal angle on a surface uniformly offset from the plasma surface by 0.25 m as in Figure \ref{lmnc_arclength_coil_pcolor}, but retaining only the 43 largest amplitude modes (the number of non-zero modes in the VMEC representation). Magnitudes of (b) $R_{m,n}^c$ and (c) $Z_{m,n}^s$ obtained by starting with the truncated single Fourier spectrum in (a). A similar dispersion to Figure \ref{lmnc_arclength_coil_pcolor} is observed for these spectra. (d) Magnitudes of the full set of $l_{m,n}^c$ (Figure \ref{lmnc_arclength_coil_pcolor}) in blue and $R_{m,n}^c$ and $Z_{m,n}^s$ computed from the truncated single Fourier series ((b) and (c)) in red, sorted in descending order.}
\label{reverse_pcolor}
\end{figure}

A spectrum which is condensed in one basis is, in general, not condensed in another given basis. We observe that the Fourier spectrum of the chosen surface is, by design, condensed in the VMEC representation, Figure \ref{RZ_pcolor}, but it is more diffuse for the various single Fourier representations, Figure \ref{l_pcolor}. To demonstrate that this is a consequence of changing basis in general rather than a weakness of the single Fourier representations specifically, we present a transformation from a condensed spectrum in a single Fourier representation to the VMEC representation in Figure \ref{reverse_pcolor}. Starting from the VMEC representation of the W7-X surface, we compute the single Fourier representation with the arclength poloidal angle on a surface uniformly offset from the plasma surface by 0.25 m. The original VMEC spectrum has a total of 43 non-zero modes while this single Fourier spectrum has 382 modes with magnitudes greater than $10^{-5}$, so we truncate this single Fourier series to only include the 43 modes with the largest magnitudes. Starting from the truncated single Fourier spectrum, we then compute the double Fourier representation in (\ref{eq:double_Fourier}) with the initial VMEC poloidal angle. We find that the VMEC spectrum calculated in this way now has a total of 734 modes with magnitudes greater than $10^{-5}$. We also present Figure \ref{mode_decay} showing the magnitudes of the single and double Fourier spectra sorted in descending order. We observe that the single Fourier series decays faster 
% and maintains smaller amplitudes
than the double Fourier representation. Thus, we can conclude that the single Fourier representation is more ``efficient,'' as fewer modes 
% with magnitude less than $10^{-5}$ 
are required to represent the surface.

	%---------------------------------

\section{Differentiation with respect to the Fourier coefficients}
\label{sec_plasma_derivs}
To take the derivative of $\chi^2_I$, where $I\in\{B,K\}$, with respect to some parameter of the plasma surface, one must consider the explicit and implicit dependence of the functional on the parameters. We will call the set of parameters of the plasma surface $\boldsymbol{\Omega}$ and a particular one $\Omega_j$. The derivative of $\chi^2_I$ with respect to $\Omega_j$ is computed using the chain rule,
\begin{equation}
    \frac{\p \chi^2_I (\boldsymbol{\Omega},\boldsymbol{\Phi}(\boldsymbol{\Omega}))}{\p \Omega_j}=\frac{\p \chi^2_I }{\p \Omega_j}\Bigg|_{\mathrm{d}\boldsymbol{\Phi}=0}+\\
    \frac{\p \chi^2_I}{\p \boldsymbol{\Phi}}\bcdot\\
    \frac{\p \boldsymbol{\Phi}}{\p \Omega_j}.
    \label{tot_chi2_dep}
\end{equation}
The bar notation here means that the first derivative on the right-hand side of (\ref{tot_chi2_dep}) is taken with $\mathrm{d}\boldsymbol{\Phi}=0$.
% , or equivalently, $\boldsymbol{\Phi}$ held constant. 
The derivatives involving $\boldsymbol{\Phi}$ indicate element-wise operation; thus, $\partial \chi^2_I/\partial \bm{\Phi}$ and $\p \bm{\Phi}/ \p \Omega_j$ are vectors of the same dimension as $\boldsymbol{\Phi}$. The dot product between these vectors in the second term of (\ref{tot_chi2_dep}) indicates contraction over $\boldsymbol{\Phi}$. The second term on the right-hand side of (\ref{tot_chi2_dep}) accounts for the implicit dependence of $\chi^2_I$ on $\Omega_j$ through $\boldsymbol{\Phi}$. In \S\ref{sec:optimal_sol} and \S\ref{sec:fixed_norm}, $\bm{\Phi}$ will gain dependence on $\bm{\Omega}$ due to constraints imposed on the total derivative.

% This term is included to allow for constraints to be imposed on the total derivative in \S\ref{sec:optimal_sol} and \S\ref{sec:fixed_norm} by giving $\bm{\Phi}$ dependence on $\bm{\Omega}$.

\subsection{Optimal Solution Constraint}
\label{sec:optimal_sol}

To take this derivative such that the current potential satisfies the linear least-squares system given in (\ref{sol_sys}), we differentiate the constraint given by (\ref{sol_sys}) with respect to the Fourier coefficients to obtain,
\begin{equation}
    \frac{\p \boldsymbol{\Phi}}{\p \Omega_j} = \mathsfbi{A}^{-1}\left(\frac{\p \boldsymbol{b}}{\p \Omega_j}-\frac{\p \mathsfbi{A}}{\p \Omega_j}\boldsymbol{\Phi}\right),
\end{equation}
which can then be substituted into (\ref{tot_chi2_dep}). Ordinarily, determining $\p \boldsymbol{\Phi} / \p \Omega_j$ would require solving a linear system involving $\mathsfbi{A}$ for each $\Omega_j$; however, this can be avoided here by using an adjoint method. The second term on the right-hand side of (\ref{tot_chi2_dep}) is,
\begin{equation}
    \frac{\p \chi^2_I}{\p \boldsymbol{\Phi}}\bcdot\\
    \mathsfbi{A}^{-1}\left[\frac{\p \boldsymbol{b}}{\p \Omega_j}-\frac{\p \mathsfbi{A}}{\p \Omega_j}\boldsymbol{\Phi}\right]=\left(\frac{\p \chi^2_I}{\p \boldsymbol{\Phi}}\right)^T\\
    \left(\mathsfbi{A}^{-1}\left[\frac{\p \boldsymbol{b}}{\p \Omega_j}-\frac{\p \mathsfbi{A}}{\p \Omega_j}\boldsymbol{\Phi}\right]\right),
    \label{eq:dot_to_transpose}
\end{equation}
where the second expression uses the transpose to perform the inner product between the vectors of the first expression as a matrix multiplication. The quantity on the left side of this product is a row vector while the quantity in parenthesis on the right is a column vector. Note that,
\begin{equation}
    \left(\frac{\p \chi^2_I}{\p \boldsymbol{\Phi}}\right)^T
    \mathsfbi{A}^{-1} = \left[\left(\mathsfbi{A}^{-1}\right)^T \frac{\p \chi^2_I}{\p \boldsymbol{\Phi}}\right]^T,
\end{equation}
meaning that the operator $\mathsfbi{A}^{-1}$ can be moved to the left side of the inner product by taking its transpose. By defining an adjoint variable $\boldsymbol{q}_I$ through,
\begin{equation}
    \mathsfbi{A}^T \boldsymbol{q}_I=\frac{\p \chi^2_I}{\p \boldsymbol{\Phi}},
    \label{adjoint_1}
\end{equation}
the inner product term \eqref{eq:dot_to_transpose} becomes
\begin{equation}
    \boldsymbol{q}_I^T \left(\frac{\p \boldsymbol{b}}{\p \Omega_j}-\frac{\p \mathsfbi{A}}{\p \Omega_j}\boldsymbol{\Phi}\right) = \boldsymbol{q}_I \bcdot \left(\frac{\p \boldsymbol{b}}{\p \Omega_j}-\frac{\p \mathsfbi{A}}{\p \Omega_j}\boldsymbol{\Phi}\right);
\end{equation}
thus, a linear system involving the matrix $\mathsfbi{A}$ only needs to be solved once for all $\Omega_j$ to get the adjoint variable since $\p \chi^2_I / \p \boldsymbol{\Phi}$ is not specific to a particular $\Omega_j$. The dependence of $\chi^2_I$ on $\boldsymbol{\Phi}$ is given by the corresponding matrices and vectors,
\begin{equation}
    \frac{\p \chi^2_I}{\p \boldsymbol{\Phi}} = 2 N_p \left( \mathsfbi{A}^I \boldsymbol{\Phi} - \boldsymbol{b}^I\right),
\end{equation}
where the definitions from Appendix A of \citep{landreman2017} have been used. Then (\ref{tot_chi2_dep}) becomes,
\begin{equation}
    \frac{\p \chi^2_I(\boldsymbol{\Omega},\boldsymbol{\Phi}(\boldsymbol{\Omega}))}{\p \Omega_j}\Bigg|_{\mathsfbi{A}\boldsymbol{\Phi}=\boldsymbol{b}}=\frac{\p \chi^2_I }{\p \Omega_j}\Bigg|_{\mathrm{d}\boldsymbol{\Phi}=0}+\boldsymbol{q}_I\bcdot\left(\frac{\p \boldsymbol{b}}{\p \Omega_j}-\frac{\p \mathsfbi{A}}{\p \Omega_j}\boldsymbol{\Phi}\right).
\end{equation}
This equation applies to both $\chi^2_B$ and $\chi^2_K$, so both adjoint variables $\boldsymbol{q}_B$ and $\boldsymbol{q}_K$ are computed from (\ref{adjoint_1}).
% with the respective $\chi^2_I$. 
Since the quantity $\chi^2_B + \lambda \chi^2_K$ is minimized with respect to $\boldsymbol{\Phi}$ by REGCOIL,
\begin{equation}
    \frac{\p \chi_B^2}{\p \boldsymbol{\Phi}} + \lambda \frac{\p \chi_K^2}{\p \boldsymbol{\Phi}} = 0.
    \label{eq:BK_relation}
\end{equation}
This means that these adjoint variables are related by the equation,
\begin{equation}
    \boldsymbol{q}_B + \lambda \boldsymbol{q}_K = 0,
    \label{eq:adjoint_variable_proportional}
\end{equation}
so it is only necessary to determine one of them by solving a linear system. 

% As stellarator optimization often requires navigating through high-dimensional $\boldsymbol{\Omega}$, the adjoint method provides a significant reduction in computational cost. 

\subsection{Fixed Norm Constraint}
\label{sec:fixed_norm}

In the previous Section, differentiation was performed at fixed $\lambda$. In order to evaluate different magnetic configurations on the same footing, we will use the freedom in $\lambda$ to fix a quantity such as $\|K\|_2$ (\ref{eq:norm_BK}) which is related to the engineering complexity of the coils. To account for this constraint, we define the following quantity,
% To add a second constraint to fix a quantity such as $\|K\|_2$, define a quantity $G$ which vanishes when the constraint is met. For $\|K\|_2$, this would be,
\begin{equation}
    G=\|K\|_2-\|K\|^{\text{target}}_2,
    \label{fixed_L2K}
\end{equation}
where $\|K\|^{\text{target}}_2$ is the fixed value desired for $\|K\|_2$. In effect, this constraint allows the coil complexity to be fixed to an acceptable level.
% This constraint is of interest since the coil complexity greatly influences whether a set of coils can be implemented. Complex coil are very expensive to construct, and there must be sufficient coil-coil separation to allow for maintenance. 
Let the optimal solution constraint be described by,
\begin{equation}
    \boldsymbol{F}=\mathsfbi{A}\boldsymbol{\Phi} - \boldsymbol{b}.
\end{equation}
Now $\lambda$ must be allowed to vary unlike in the previous Section. By taking the total differentials of $\boldsymbol{F}$ and $G$, we obtain,
\begin{equation}
    \begin{split}
        \mathrm{d}\boldsymbol{F} &= \sum_j \left( \frac{\p \mathsfbi{A}}{\p \Omega_j}\boldsymbol{\Phi} - \frac{\p \boldsymbol{b}}{\p \Omega_j} \right)\mathrm{d}\Omega_j + \mathsfbi{A} \mathrm{d}\boldsymbol{\Phi} +\left( \mathsfbi{A}^K \boldsymbol{\Phi} - \boldsymbol{b}^K \right)\mathrm{d}\lambda = 0\\
        \mathrm{d}G &= \sum_j \frac{\p G}{\p \Omega_j}\Bigg|_{\mathrm{d}\boldsymbol{\Phi}=0} \mathrm{d}\Omega_j + \frac{\p G}{\p \boldsymbol{\Phi}} \cdot \mathrm{d}\boldsymbol{\Phi} = 0.
    \end{split}
    \label{eq:constraint_differentials}
\end{equation}
The differential of $\lambda$ can be eliminated from these equations, as described in \S\ref{app:fixed_norm}, and $\p \boldsymbol{\Phi} / \p \Omega_j$ subject to the constraints $\boldsymbol{F}=0$ and $G=0$ is obtained,
\begin{equation}
\begin{split}
    \frac{\p \boldsymbol{\Phi} }{\p \Omega_j}\Bigg|_{\boldsymbol{F}=0,\ G=0} &= -\mathsfbi{A}^{-1}\Bigg[ \left( \frac{\p \mathsfbi{A}}{\p \Omega_j}\boldsymbol{\Phi} - \frac{\p \boldsymbol{b}}{\p \Omega_j} \right)\\
    &+ \frac{(\mathsfbi{A}^K\boldsymbol{\Phi} - \boldsymbol{b}^K)}{\widetilde{\boldsymbol{q}}\bcdot(\mathsfbi{A}^K\boldsymbol{\Phi} - \boldsymbol{b}^K)} \left[ \frac{\p G}{\p \Omega_j}\Bigg|_{\mathrm{d}\boldsymbol{\Phi}=0} - \widetilde{\boldsymbol{q}}\cdot \left( \frac{\p \mathsfbi{A}}{\p \Omega_j}\boldsymbol{\Phi} - \frac{\p \boldsymbol{b}}{\p \Omega_j} \right) \right] \Bigg],
\end{split}
\label{eq:fixed_norm_implicit}
\end{equation}
where an additional adjoint variable $\widetilde{\bm{q}}$ is chosen to satisfy,
\begin{equation}
    \mathsfbi{A}^T \widetilde{\boldsymbol{q}}=\frac{\p G}{\p \boldsymbol{\Phi}}.
    \label{eq:fixed_norm_adjoint}
\end{equation}
For the case of $G$ given by (\ref{fixed_L2K}), the explicit derivative $\p G / \p \Omega_j = 0$, and $\p G /\p \boldsymbol{\Phi}$ is proportional to $\p \chi_K^2 /\p \boldsymbol{\Phi}$. These conditions cause the implicit dependence in (\ref{tot_chi2_dep}) to cancel, leaving only the explicit dependence,
\begin{equation}
    \frac{\p \chi^2_I(\boldsymbol{\Omega},\boldsymbol{\Phi}(\boldsymbol{\Omega}))}{\p \Omega_j}\Bigg|_{ \mathsfbi{A}\boldsymbol{\Phi}=\boldsymbol{b}, \ \|K\|_2=\|K\|^{\text{target}}_2 }=\frac{\p \chi^2_I }{\p \Omega_j}\Bigg|_{\mathrm{d}\boldsymbol{\Phi}=0},
\end{equation}
as discussed in \S\ref{app:fixed_norm_and_offset}.

\subsection{Uniform Offset Constraint}
\label{sec:offset}

Up to this point, the derivatives have been computed with a fixed winding surface. Without placing an additional constraint on the winding surface, the plasma boundary may move uniformly toward the winding surface in order to minimize the coil-plasma distance, as the shaping components of the magnetic field decay with distance from the coils \citep{Boozer2000,landreman2016}. For this reason, we instead compute derivatives with respect to the plasma surface parameters while maintaining the constraint that the coil winding surface is uniformly offset a distance $a$ from the plasma surface. The coil surface is then defined by,
\begin{equation}
    \boldsymbol{r}'=\boldsymbol{r}+a\boldsymbol{\hat{n}}.
    \label{eq:coil_offset}
\end{equation}
The coil offset distance must similarly be constrained by the principal curvatures as,
\begin{align}
0 <  a < \frac{1}{|\min\left\{\kappa_1, \kappa_2\right\}|}.
    % |a| < \min\left\{\frac{-\text{sign}(a)}{\kappa_1}, \frac{-\text{sign}(a)}{\kappa_2}\right\} .
\label{eq:max_offset_coil}
\end{align}
Unlike the restriction on the offset surface that defines the poloidal angle \eqref{eq:angle_max_offset}, the coil offset distance is constrained by \eqref{eq:max_offset_coil} to be greater than zero since the coils must lie in the vacuum region. We remark that as the plasma surface evolves during the optimization, a constraint must be placed on the principal curvatures such that the above inequality is satisfied. As concavity of the plasma boundary has been shown to be correlated with coil complexity \citep{Paul2018}, placing such a constraint on the curvature of the boundary is reasonable and can be performed with a penalty formulation \citep{Paul2020b}. If desired, the winding surface could be offset by a non-uniform function on the surface, and instead, constraints would be imposed on the minimum and maximum offset distance.

This constraint couples the geometry of the winding surface to that of the plasma surface and, therefore, introduces additional derivatives with respect to the plasma surface Fourier amplitudes. For example, $\mathsfbi{A}$ gains additional dependence through the $\mathsfbi{A}^K$ term \eqref{eq:A_b}. Some equations are given in \S\ref{app:fixed_offset} showing how the derivatives of the coil winding surface position and normal vectors become coupled to the plasma surface through (\ref{eq:coil_offset}).

No new adjoint variables are needed to impose this constraint. With the combination of this and the fixed $\|K\|_2$ constraint, $G$ given by (\ref{fixed_L2K}) now has explicit dependence on the plasma surface parameters; however, $\p G /\p \boldsymbol{\Phi}$ is still proportional to $\p \chi_K^2 /\p \boldsymbol{\Phi}$. The total derivatives of $\chi^2_B$ and $\chi^2_K$ are then given by,
%The modified expressions for the derivatives are not included here for brevity.
\begin{equation}
\begin{split}
    \frac{\p \chi^2_B(\boldsymbol{\Omega},\boldsymbol{\Phi}(\boldsymbol{\Omega}))}{\p \Omega_j}\Bigg|_{ \mathsfbi{A}\boldsymbol{\Phi}=\boldsymbol{b}, \ \|K\|_2=\|K\|^{\text{target}}_2, \ \boldsymbol{r}'=\boldsymbol{r}+a\boldsymbol{\hat{n}} }&=\frac{\p \chi^2_B }{\p \Omega_j}\Bigg|_{\mathrm{d}\boldsymbol{\Phi}=0}\\
    &-\lambda \left( \frac{\chi^2_K}{a_{\text{coil}}}\frac{\p a_{\text{coil}}}{\p \Omega_j} - \frac{\p \chi^2_K }{\p \Omega_j}\Bigg|_{\mathrm{d}\boldsymbol{\Phi}=0} \right)\\
    \frac{\p \chi^2_K(\boldsymbol{\Omega},\boldsymbol{\Phi}(\boldsymbol{\Omega}))}{\p \Omega_j}\Bigg|_{ \mathsfbi{A}\boldsymbol{\Phi}=\boldsymbol{b}, \ \|K\|_2=\|K\|^{\text{target}}_2, \ \boldsymbol{r}'=\boldsymbol{r}+a\boldsymbol{\hat{n}} }&=\frac{\chi^2_K}{a_{\text{coil}}}\frac{\p a_{\text{coil}}}{\p \Omega_j}.
\end{split}
\end{equation}
A detailed derivation of these equations is given in \S\ref{app:fixed_norm_and_offset}. Note that the implicit dependence of $\chi^2_B$ is $-\lambda$ times the implicit dependence of $\chi^2_K$, and the total dependence of $\chi^2_K$ is such that $\|K\|_2$ vanishes.

% , though the values of the derivatives will change.

% Previously, some quantities only depended on the coil surface, meaning their derivatives with respect to the plasma Fourier amplitudes vanish; however, for this constraint, these quantities now depend on these amplitudes. 
%Other quantities also pick up additional dependence on the amplitudes through the coil surface. 

% This sets a maximum offset distance. At this distance, there will be points on the offset surface where the curvature diverges, and beyond this distance, the offset surface will intersect itself. 

Since higher-order derivatives of the position vector are needed to compute the normal vector and its derivatives, any high-frequency noise in the plasma surface representation is amplified when taking derivatives with the constraint given by \eqref{eq:coil_offset}. For this reason, the poloidal angle defined with respect to the arclength on the winding surface as described in \S \ref{sec:poloidal_angles} greatly improves the numerical performance with this constraint.

\section{Benchmarks and demonstrations}
\label{sec_benchmarks}
In this Section, we present several benchmark calculations to verify the derivatives and constraints from the previous Section.
\subsection{Finite-difference derivatives}
\begin{figure}	%---------------------------------
\centering 
\begin{subfigure}[b]{0.49\textwidth}
\hspace*{0cm}\includegraphics[trim=0cm 0cm 7cm 3.5cm, clip, width=1.0\textwidth]{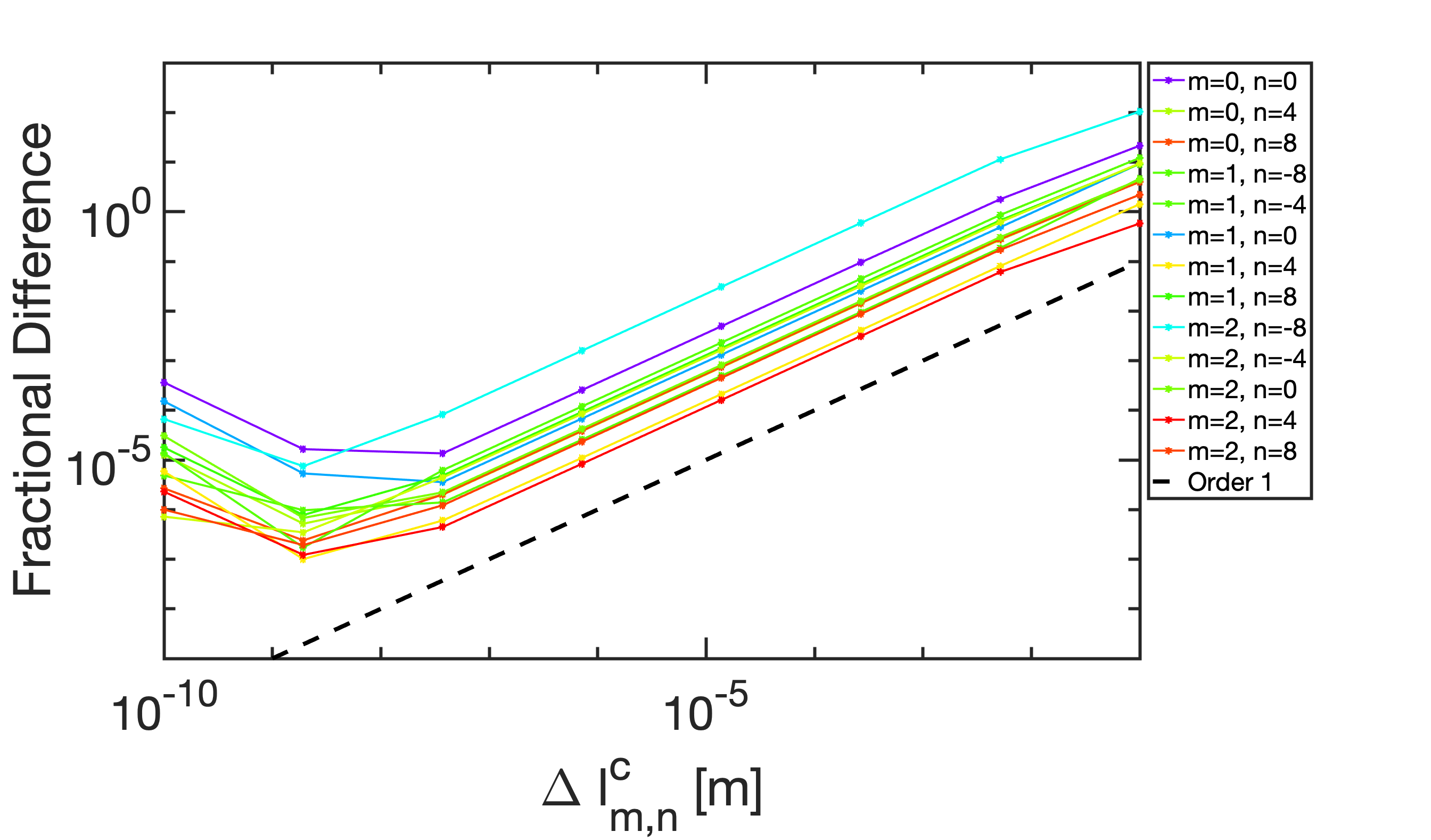}
\caption{}
% \caption{Fractional difference between analytic and numeric $\partial|| \textbf{B}||_2/\partial l_{m,n}^c$ with uniform offset.}
\label{BL2}
\end{subfigure}
\begin{subfigure}[b]{0.49\textwidth}
\hspace*{0cm}\includegraphics[trim=0cm 0cm 7cm 3.5cm, clip, width=1.0\textwidth]{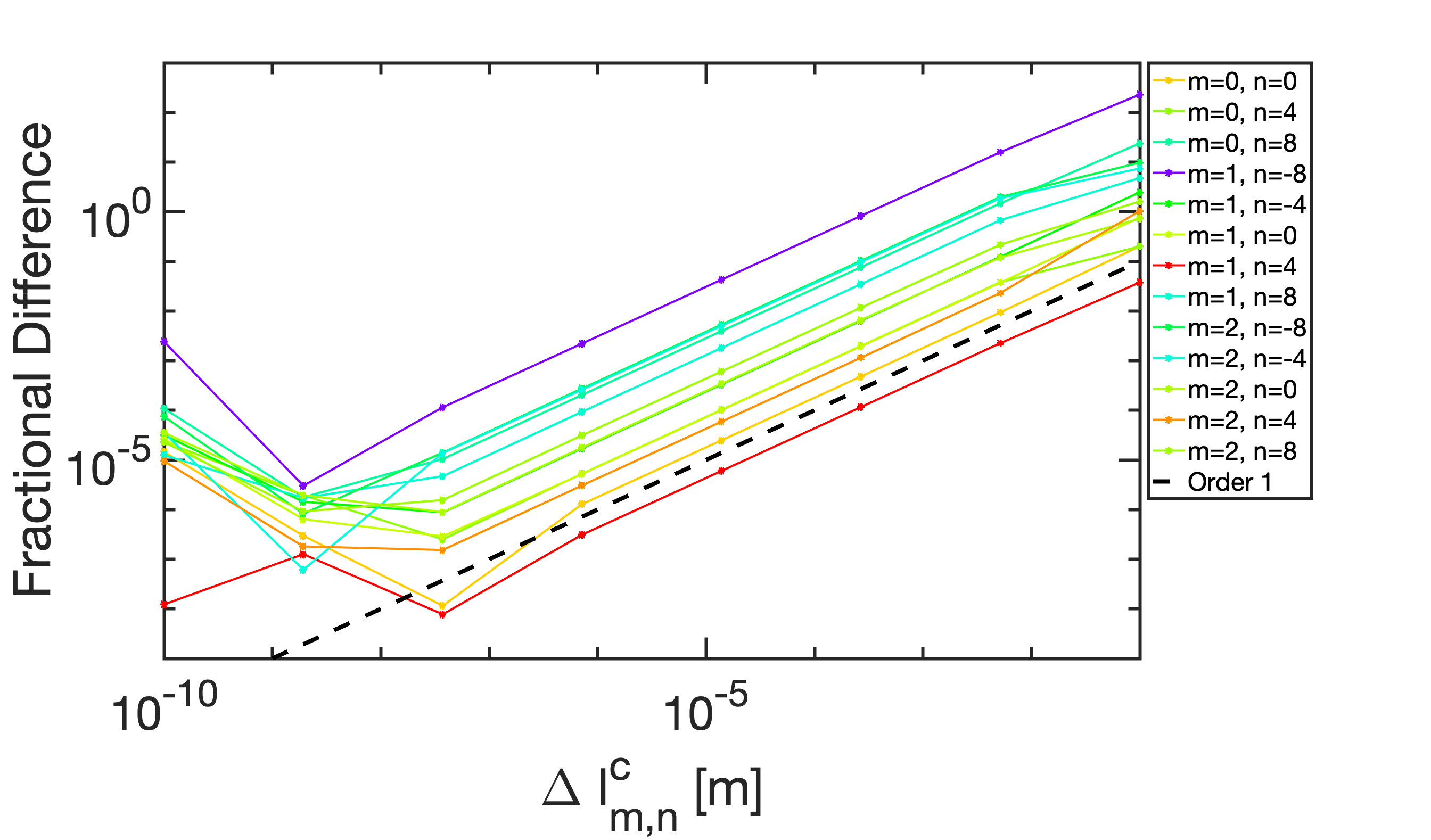}
\caption{}
% \caption{Fractional difference between analytic and numeric $\partial|| \textbf{K}||_2/\partial l_{m,n}^c$ with uniform offset.}
\label{KL2}
\end{subfigure}
\begin{subfigure}[b]{0.49\textwidth}
\includegraphics[trim=0cm 0cm 7cm 3.5cm, clip, width=1.0\textwidth]{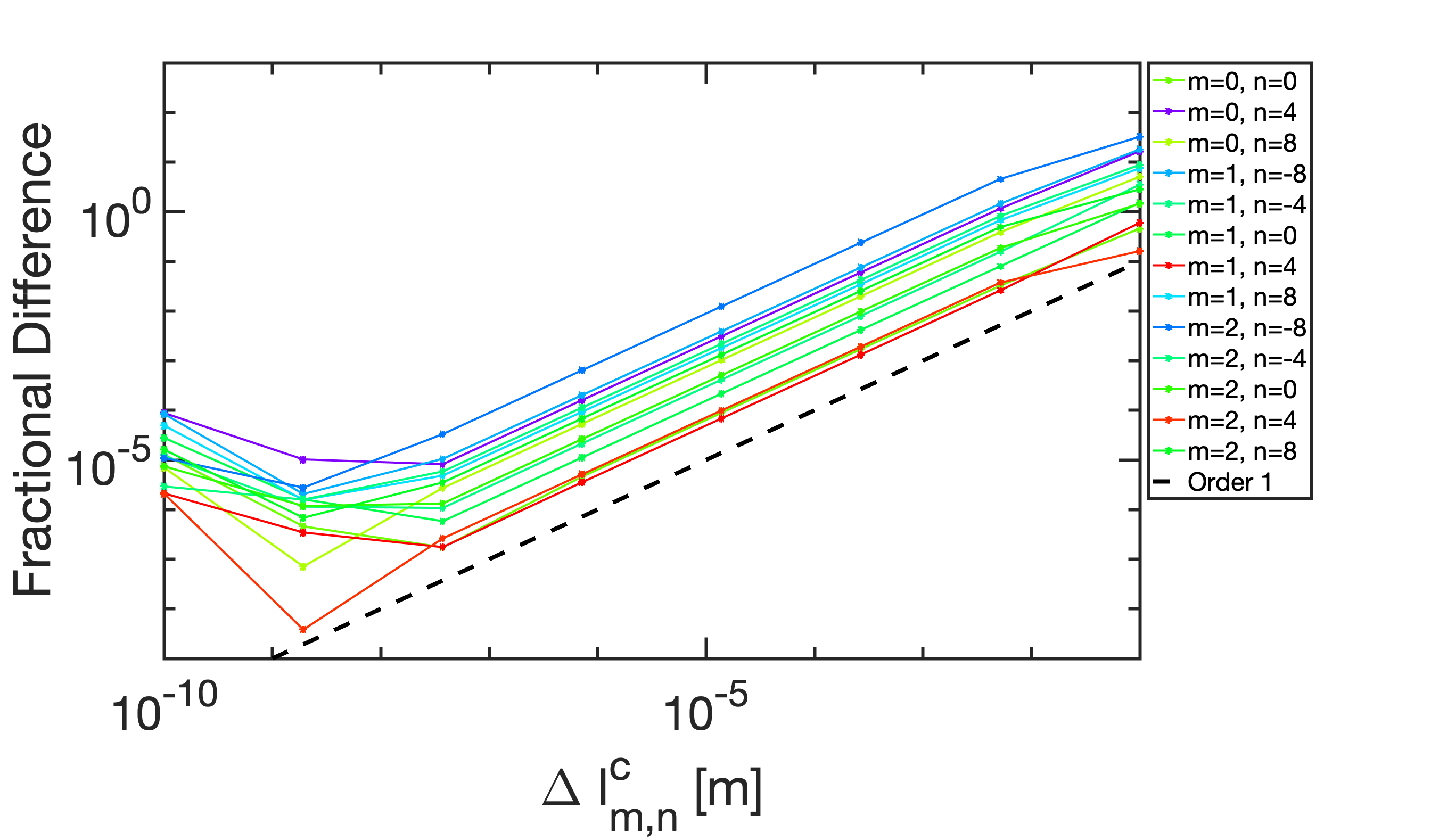}
\caption{}
\label{BL2_fixed_norm}
\end{subfigure}
\caption{Modes with $m$ up to 2 and $n$ up to $\pm 2 N_p$ are displayed with lines on the red end of the spectrum corresponding to larger magnitudes of the derivative and lines on the blue end of the spectrum corresponding to smaller magnitudes. We present the fractional difference between analytic and numerical calculations of (a) $\p \|B_n\|_2 / \p l^c_{m,n} $, (b) $\p \|K\|_2 / \p l^c_{m,n} $, and (c) $\p \|B_n\| / \p l^c_{m,n} $ while holding $\|K\|_2$ fixed. For each of these, the offset distance is additionally fixed.
% fixed for modes with $m$ up to 3 and $n$ up to $\pm 3 N_p$. 
% Lines on the red end of the spectrum correspond to larger values of $\p \|B_n\|_2 / \p l^c_{m,n} $ while lines on the blue end of the spectrum correspond to larger values to $\p \|B_n\|_2 / \p l^c_{m,n} $.
}
\end{figure}	%---------------------------------
The derivatives of $\chi^2_I$ are implemented analytically as described in the previous Section rather than with a finite-difference approximation. The former is advantageous over the latter in that it takes far less computation time for a high-dimensional derivative; however, the finite-difference result should converge to the analytic result as the step size is reduced. This is used to check the accuracy of the analytic derivatives. Consider a numerical forward-difference approximation of $\p \chi^2_I / \p \Omega_j$,
\begin{equation}
    \left(\frac{\p \chi^2_{I}}{\p \Omega_j}\right)_{\text{num.}}=\frac{\chi^2_I(\Omega_j+\Delta\Omega)-\chi^2_I(\Omega_j)}{\Delta\Omega}.
\end{equation}
% This means that the first term in the numerator of the right hand side is perturbed by $\Delta \Omega$ only in the $\Omega_j$ parameter, while the other parameters are left unchanged. 
When doing this calculation, 
% it is important to account from 
the perturbation of the current potential corresponding to the perturbation of $\Omega_j$ is included in order to account for the implicit dependence of $\chi^2_I$ on the plasma surface. This expression is a first-order approximation, so it should agree with the analytic expression to $\textit{O}(\Delta \Omega)$. 

We now define the fractional difference between the analytic and numerical derivatives as
\begin{equation}
    \text{fractional difference}=\left(\frac{\p \chi^2_I}{\p \Omega_j} - \left(\frac{\p \chi^2_{I}}{\p \Omega_j}\right)_{\text{num.}}\right) \left(\frac{\p \chi^2_I}{\p \Omega_j}\right)^{-1}.
\end{equation}
The following results were generated using the QHS46 equilibrium \citep{qhs46} with a $0.25\ \text{m}$ uniformly offset coil winding surface which was held fixed during the calculation of derivatives with respect to the plasma parameters. A plot of the fractional difference for the derivatives of $\|B\|_2$ with only the optimal solution constraint and uniform offset constraint imposed is given in Figure \ref{BL2}.
% for small $m$ and $n$ 
The surface is described by the single Fourier representation given in (\ref{l_series_arclength}) with the arclength angle  defined on a surface chosen to have the same $0.25\ \text{m}$ uniform offset as the coil winding surface. A similar fractional difference plot is given for $\|K\|_2$ in Figure \ref{KL2}. 
% With this representation, the expansion parameters $\Omega_j$ are the Fourier coefficients $l_{m,n}^c$, where $m$ and $n$ are two integers which index the modes, and $c$ indicates that they are the coefficients for the cosine functions. 
% The sine coefficients are not needed to describe a stellarator symmetric configuration. 
We find that the fractional difference scales linearly with $\Delta \Omega$ for $10^{-7} \lesssim \Delta \Omega \lesssim 10^{-2}$.
% specifically the slopes match that of the dotted line indicating that the fractional difference is first order in $\Delta \Omega$, as we would expect from this forward difference approximation of the derivative. 
For $\Delta \Omega \lesssim 10^{-7}$, the round-off error begins to dominate, and the linear relationship is no longer observed \citep{Sauer2012}. Similar trends are also observed in the fractional difference for $\|B\|_2$ with the addition of the fixed $\|K\|_2$ constraint, presented in Figure \ref{BL2_fixed_norm}. For this calculation with the fixed $\|K\|_2$ constraint,  $\|K\|_2^{\text{target}}=1.6\ \text{MA m}^{-1}$ was used.
% \begin{figure}
% \centering 
% \hspace*{0cm}\includegraphics[trim=5cm 0cm 9cm 3.5cm, clip, width=1.0\textwidth]{frac_diff_qhs46_L2B_fixed_norm}
% \caption{Plot of fractional difference between analytic and numerical calculations of $\p \|B_n\| / \p l^c_{m,n} $ while holding $\|K\|_2$ fixed for modes with $m$ up to 3 and $n$ up to $\pm 3 N_p$. Lines on the red end of the spectrum correspond to larger values of $\p \|B_n\|_2 / \p l^c_{m,n} $ while lines on the blue end of the spectrum correspond to larger values to $\p \|B_n\|_2 / \p l^c_{m,n} $.}
% \label{BL2_fixed_norm}
% \end{figure}

%---------------------------------
\subsection{Area shape gradient benchmark}
\label{sec:area_shape_gradient}
\begin{figure}	%---------------------------------
\centering 
\begin{subfigure}[b]{0.49\textwidth}
\hspace*{0cm}
\includegraphics[width=1.0\textwidth]{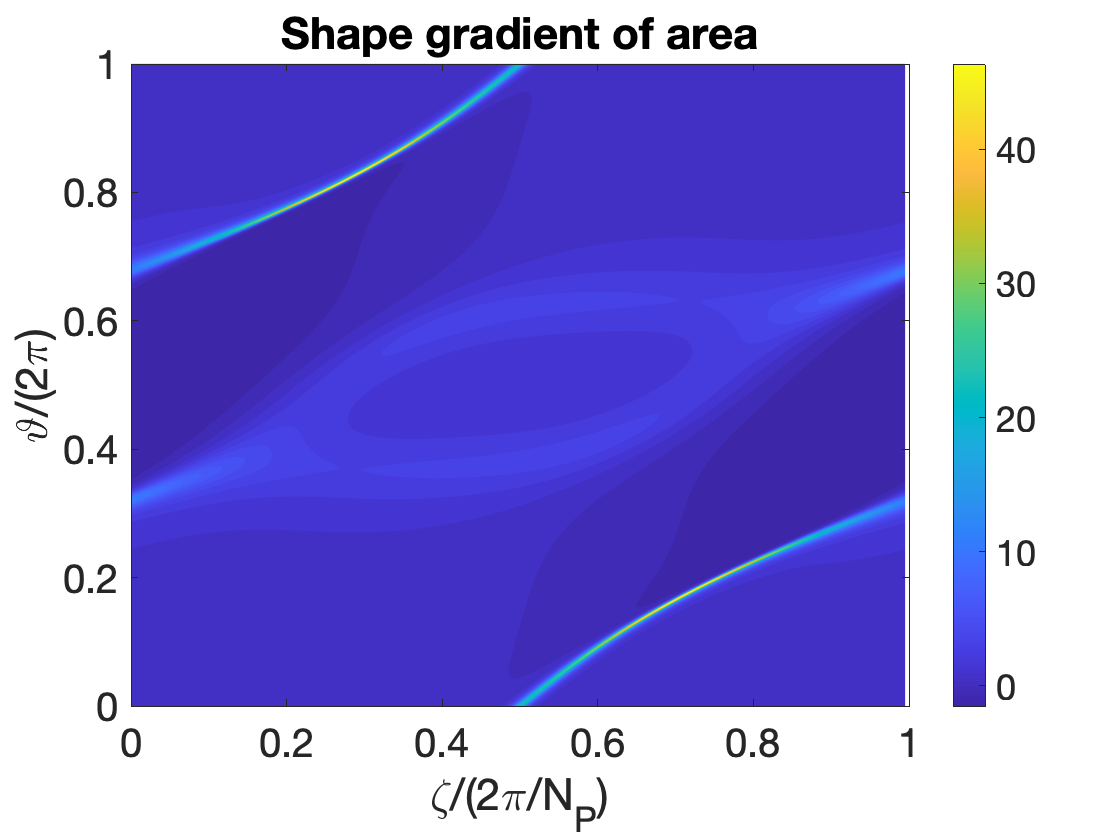}
\caption{}
\label{shape_grad_area_calc}
\end{subfigure}
\begin{subfigure}[b]{0.49\textwidth}
\hspace*{0cm}
\includegraphics[width=1.0\textwidth]{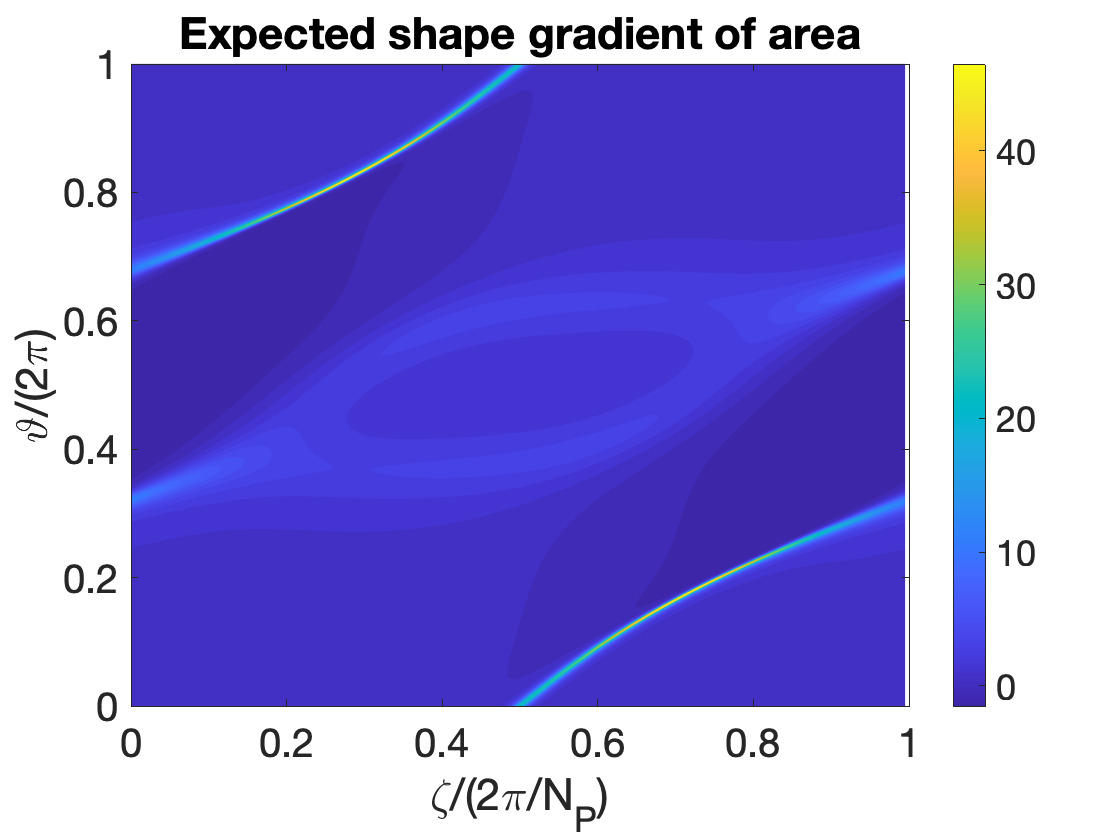}
\caption{}
\label{shape_grad_area_exp}
\end{subfigure}
\begin{subfigure}[b]{0.49\textwidth}
\centering
\includegraphics[trim=3cm 3cm 4cm 6cm,clip,width=1.0\textwidth]{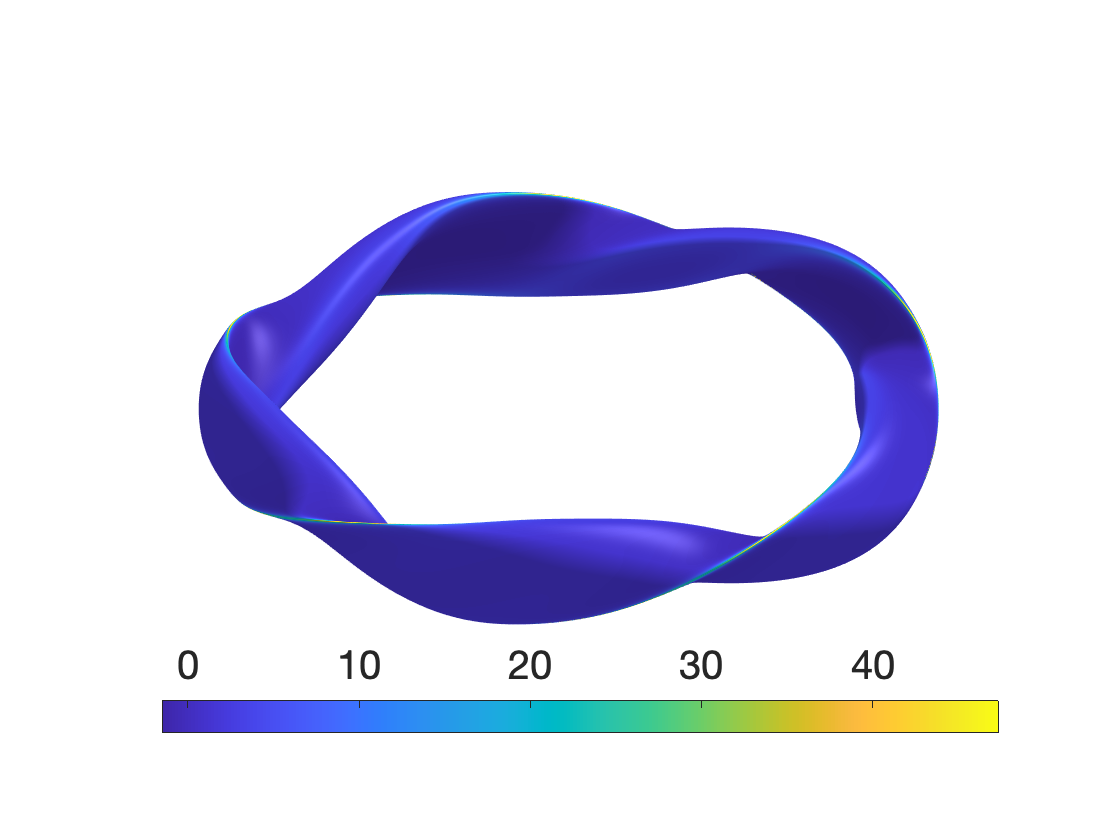}
\caption{}
\end{subfigure}
\begin{subfigure}[b]{0.49\textwidth}
\centering
\includegraphics[trim=0cm 0cm 2cm 1cm,clip,width=1.0\textwidth]{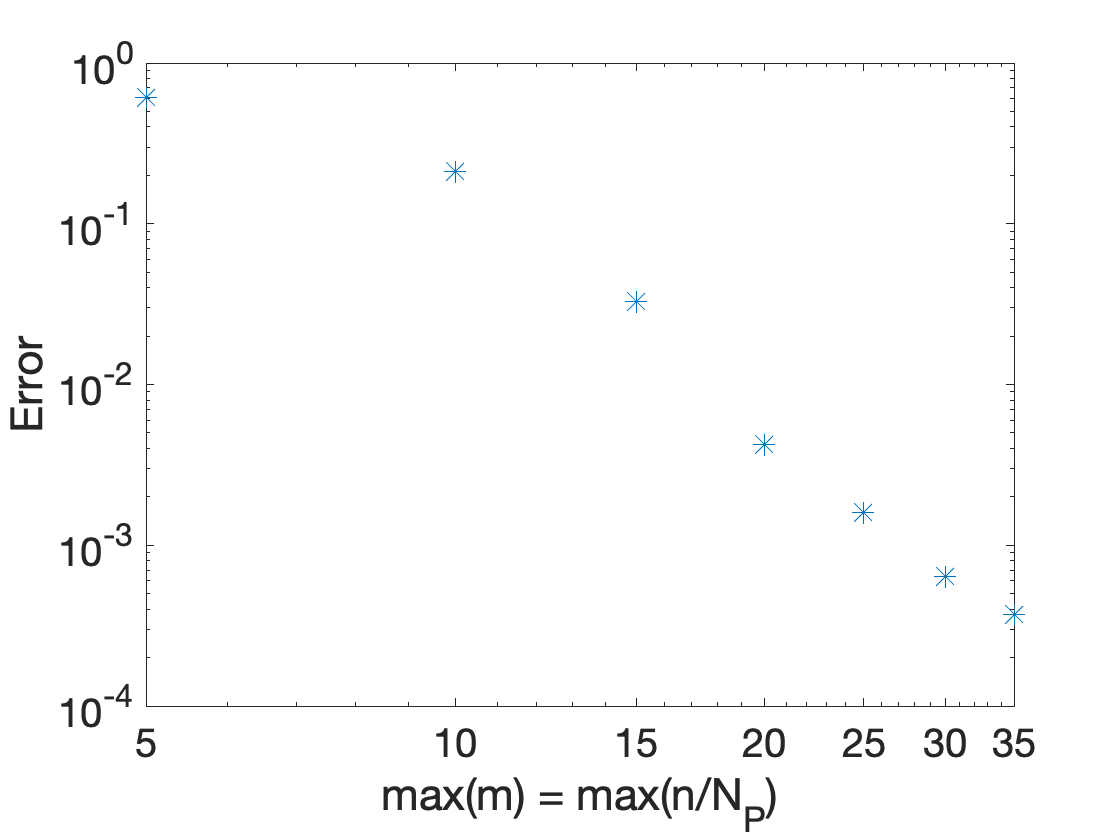}
\caption{}
\label{fig:resolution_scan}
\end{subfigure}
\caption{Plot in the $\vartheta$ and $\zeta$ plane of (a) the shape gradient of the plasma area calculated in the REGCOIL code using a solution of (\ref{shape_grad_lin_sys}) for a square system with $m \le 35$ and $|n/N_P| \le 35$ and (b) the shape gradient of the plasma area calculated using the explicit form involving the mean curvature \eqref{eq:mean_curvature}. In (c), the same quantity plotted in (a) is displayed on the W7-X plasma boundary, and in (d) the error defined in \eqref{eq:area_error} is shown as a function of the maximum poloidal and toroidal mode number used to compute the shape gradient.}
% . (b) Plot in the $\theta$ and $\zeta$ plane of}
\end{figure}

% \subsection{Mean Curvature}
As a benchmark for calculating shape gradients with the modified REGCOIL code, the shape gradient of the plasma surface area is computed. There is an analytic expression for this shape gradient \citep{Landreman2018}, which is given by,
\begin{equation}
    S_{a_{\text{plasma}}}=2H,
    \label{eq:mean_curvature}
\end{equation}
where $H = \frac{1}{2}(\kappa_1 + \kappa_2)$ is the mean curvature of the plasma surface. Again, we assume the convention that positive $H$ indicates convexity. The shape gradient of the plasma area calculated using \eqref{shape_grad_lin_sys} is given in Figure \ref{shape_grad_area_calc}. The calculation is performed for the W7-X boundary. (This is the boundary of a fixed-boundary equilibria that preceded the coil design and does not include coil ripple.) The shape gradient is computed on a grid of $N_{\overline{\theta}}=200$ grid points in the poloidal angle and $N_{\zeta} = 200$ grid points in the toroidal angle. The derivatives of the area are computed for $m \le 35$ and $|n/N_P| \le 35$, and the shape gradient is discretized with a Fourier series with the same set of modes, so \eqref{shape_grad_lin_sys} is a square linear system. We compare this to the expected result,
% calculated from the mean curvature, 
given in Figure \ref{shape_grad_area_exp}. 
% The results are presented in a two dimensional plane of the angles $\overline{\theta}$ and $\zeta$ used to parameterize the plasma surface.
% These plots use 192 values of $\theta$ and 192 values of $\zeta$ in each toroidal period. 
% The mode number $m$ was chosen from integers in range $[0,64]$ and the mode number $n$ from integer multiples of 4 ($N_p=5$ for this plasma surface) in the range $[-4 \cdot 64,4 \cdot 64]$. These modes were also filtered to only include those with $l^c_{m,n} \ge 10^{-7}\ \text{m}$. 
The average error in the shape gradient,
%The root mean-squared average of the difference between these plots defined by,
\begin{equation}
    % \text{rms} = \sqrt{\frac{1}{4\pi^2}\int\left( S_{a_{\text{plasma}}}+2H_{\text{plasma}} \right)^2 \mathrm{d}\theta\  \mathrm{d}\zeta},
    \text{error} = \frac{\int_{S_{\text{plasma}}} \, |S_{a_{\text{plasma}}} - 2 H| \ \mathrm{d}a }{\int_{S_{\text{plasma}}} \, |2 H| \ \mathrm{d}a},
    \label{eq:area_error}
\end{equation}
% which is evaluated on a discrete grid of $\overline{\theta}$ and $\zeta$, 
is computed to be $3.69 \times 10^{-4}$. In Figure \ref{fig:resolution_scan}, we display the convergence of the error with respect to the modes retained in the derivatives and in computing the shape gradient. We see that the expected and computed shape gradients converge to each other with increasing numerical resolution.
%gives $0.21\ \text{m}^{-1}$. 

% This integral is dominated by the regions of maximum shape gradient, shown in yellow in figure \ref{shape_grad_area_calc} and figure \ref{shape_grad_area_exp}, so the order of magnitude of the shape gradient in these regions is $50\ \text{m}^{-1}$; 
	%---------------------------------

\subsection{Shape gradients of $\|B_n\|_2$ and $\|K\|_2$}
\label{sec:shape_gradient_metrics}

In this Section, (\ref{shape_grad_lin_sys}) is used to obtain the shape gradients of $\|B_n\|_2$ and $\|K\|_2$ for the QHS46 and W7-X configurations using the derivatives obtained from the adjoint method. For both configurations, we choose the winding surface to be uniformly offset from the plasma boundary \eqref{eq:coil_offset}. In the case of W7-X we choose $a = 0.37$ m to match the minimum coil-plasma distance of the actual W7-X winding surface. For QSH46 we choose $a = 0.21$ m such that the offset distance is scaled according to the minor radius of the plasma boundary. For the QHS46 calculations, a fixed value of $\|K\|_2^{\text{target}}=1.7\ \text{MA m}^{-1}$ was used while for W7-X, it was taken to be ${2.4}\text{ MA m}^{-1}$. The computation of the derivatives necessary for the evaluation of these shape gradients required about 7 hours on a single AMD Opteron 6136 CPU. The cost of construction of the shape gradient from the parameter derivatives is negligible in comparison (< 1 minute on a laptop).

\begin{figure}	%---------------------------------
\centering 
\begin{subfigure}[b]{0.49\textwidth}
\hspace*{0cm}\includegraphics[trim=1cm 1cm 1cm 1cm, clip,width=1.0\textwidth]{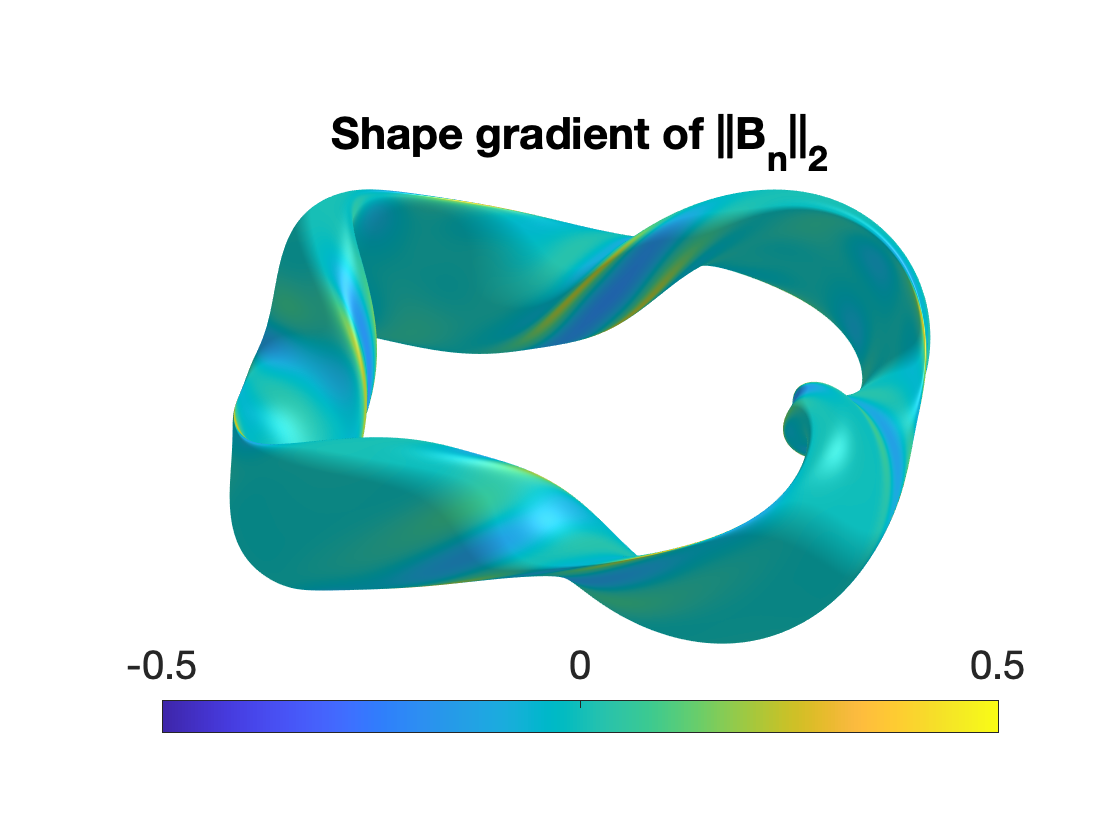}
\caption{}
\label{shape_grad_normB}
\end{subfigure}
\begin{subfigure}[b]{0.49\textwidth}
\hspace*{0cm}\includegraphics[trim=1cm 1cm 1cm 1cm, clip,width=1.0\textwidth]{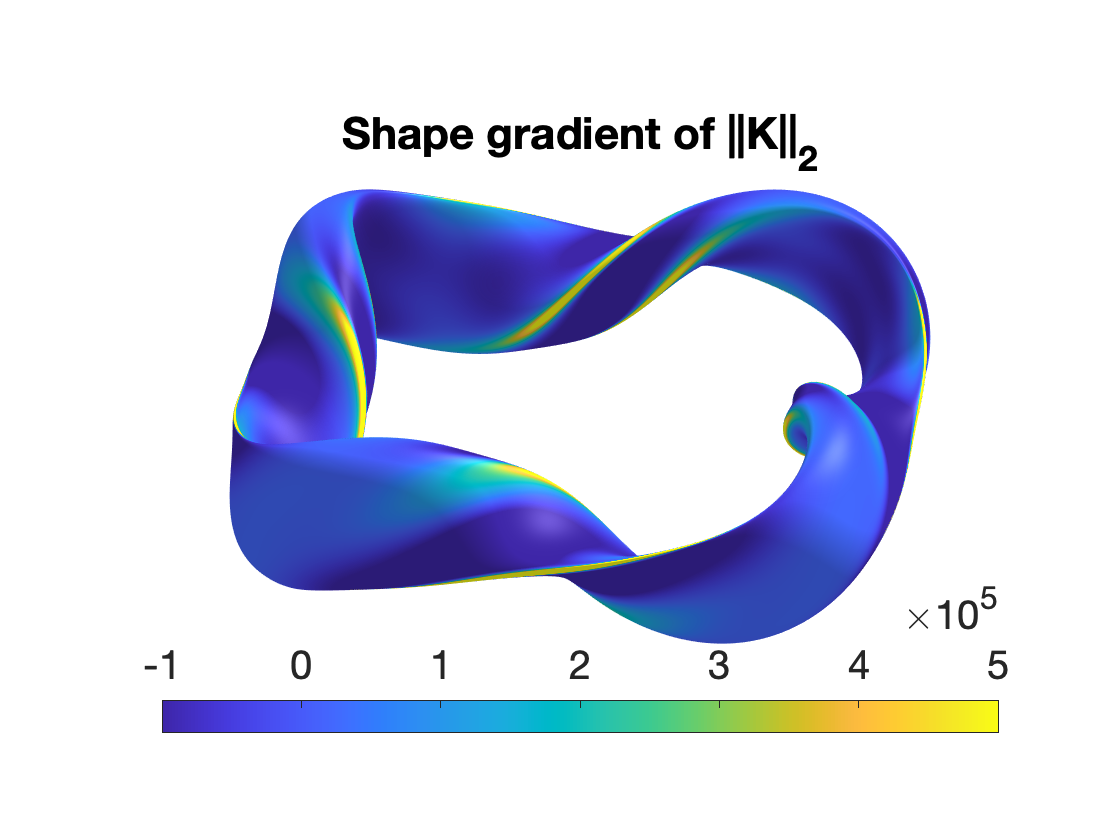}
\caption{}
\label{shape_grad_normK}
\end{subfigure}
\begin{subfigure}[b]{0.49\textwidth}
\hspace*{0cm}\includegraphics[trim=1cm 1cm 1cm 1cm, clip,width=1.0\textwidth]{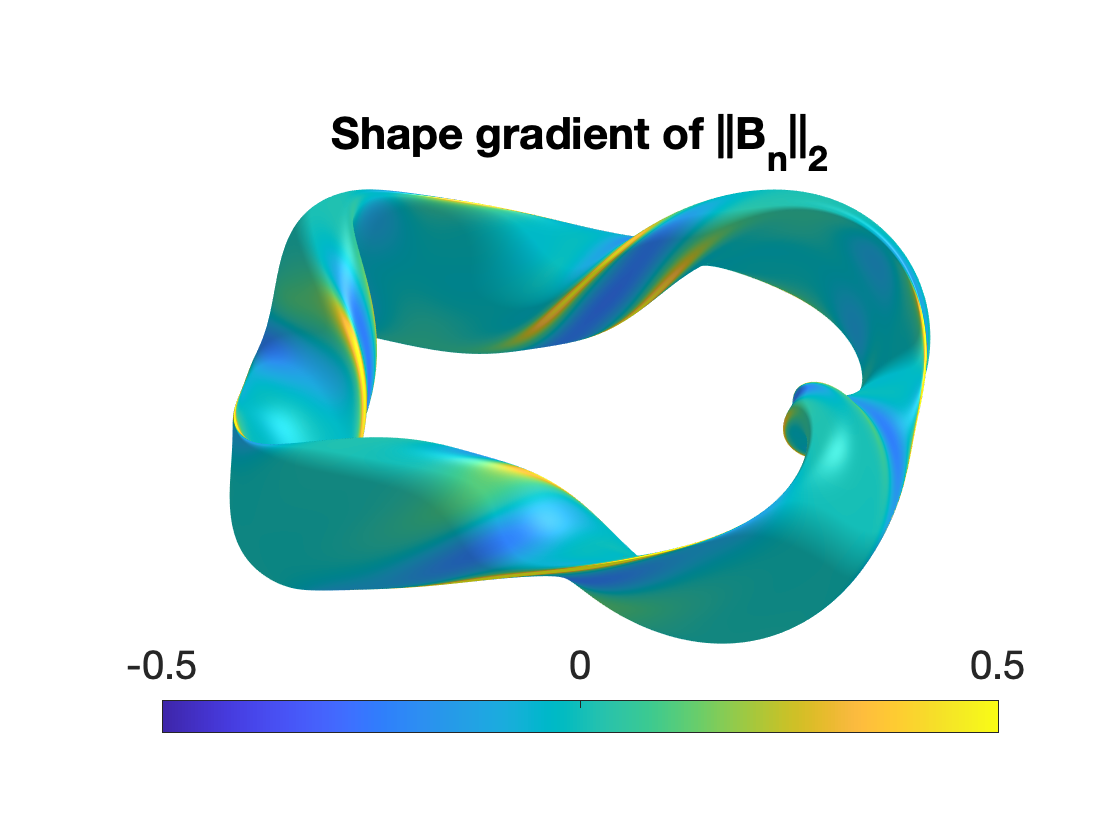}
\caption{}
\label{shape_grad_normB_fixed}
\end{subfigure}
\begin{subfigure}[b]{0.49\textwidth}
\hspace*{0cm}\includegraphics[trim=1cm 1cm 1cm 1cm, clip,width=1.0\textwidth]{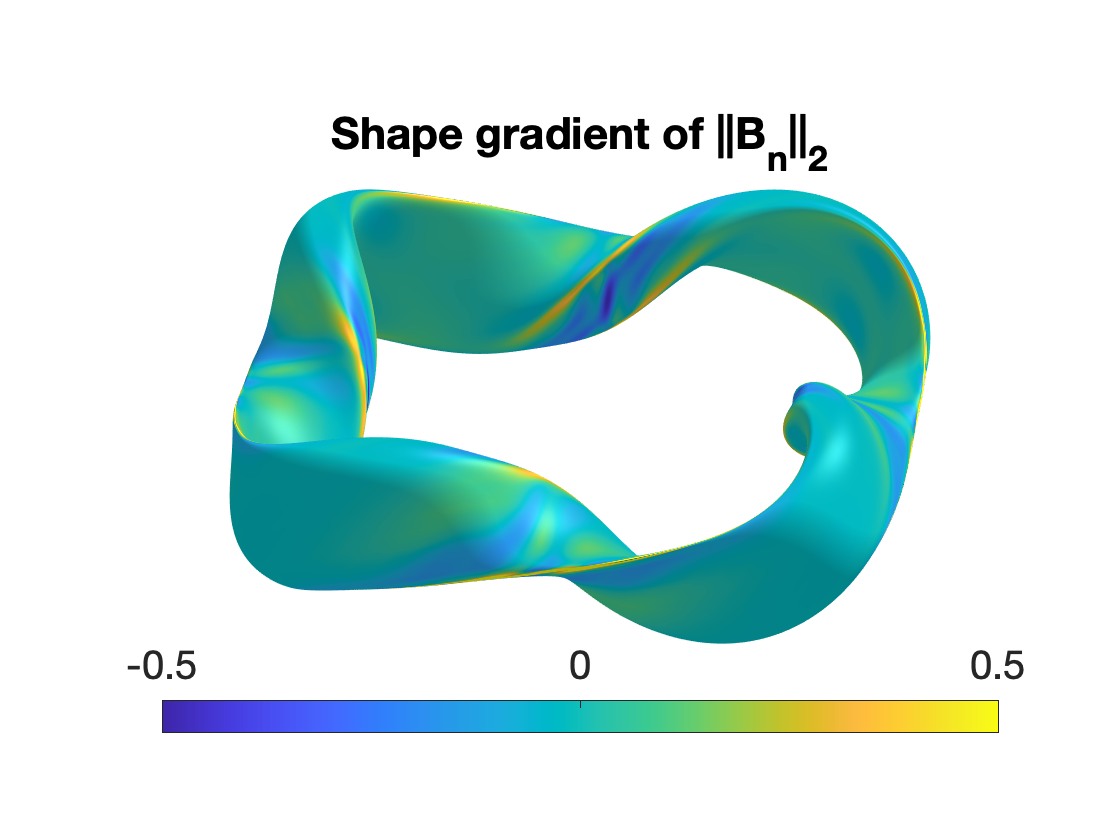}
\caption{}
\label{shape_grad_normB_fixed_surface}
\end{subfigure}
\caption{The shape gradient of (a) $\|B_n\|_2$, (b) $\|K\|_2$, (c) $\|B_n\|_2$ with fixed $\|K\|_2$, and (d) $\|B_n\|_2$ with fixed $\|K\|_2$ and fixed offset distance is computed for the QHS46 boundary with a uniform offset winding surface of 0.21 m. The shape gradient is computed using a linear solve of (\ref{shape_grad_lin_sys}) for a square system with modes $m \le 30$ and $|n/N_P| \le 30$.
% \textcolor{red}{Discussion of range of colorscales or max/min values}
% on a 3D mapping of the plasma surface 
% of the shape gradient of  calculated 
% in the REGCOIL code
% using a  (b) Plot on a 3D mapping of the plasma surface of the shape gradient of $\|K\|_2$ calculated in the REGCOIL code using a linear solve of Eq.~(\ref{shape_grad_lin_sys}) for a square system. (c) Plot on a 3D mapping of the plasma surface of the shape gradient of $\|B_n\|_2$ with fixed $\|K\|_2$ calculated in the REGCOIL code using a linear solve of Eq.~(\ref{shape_grad_lin_sys}) for a square system.
}
\label{fig:qhs46_sg}
\end{figure}

In Figure \ref{fig:qhs46_sg}, results are shown on the three dimensional shape of the QHS46 plasma surface. The parameter derivatives with respect to modes with $m \le 30$ and $|n/N_P| \le 30$ are computed on a grid with $N_{\overline{\theta}} = N_{\zeta} = 100$, and the shape gradient is discretized with a Fourier series with the same set of modes. 
The shape gradient of $\|B_n\|_2$ for only the optimal solution constraint is shown in Figure \ref{shape_grad_normB} and similarly for $\|K\|_2$ in Figure \ref{shape_grad_normK}. We see that for both of these quantities, the shape gradient is large and positive in the regions of convex curvature and strongly negative in the nearby regions where the surface curvature is locally reduced. This indicates that in order to reduce both of these figures of merit, the sharp ridge features on the boundary must be locally rounded. Similar features are present when the shape gradient of $\|B_n\|_2$ is computed with the fixed $\|K\|_2$ constraint as shown in Figure \ref{shape_grad_normB_fixed}, though the magnitude is slightly increased. In Figure \ref{shape_grad_normB_fixed_surface}, the shape gradient with the additional uniform offset constraint is qualitatively similar, but an additional feature is present near the triangle-shaped cross-section.

% In Figure \ref{shape_grad_normB}, for example, deforming the yellow regions of the plasma surface inward or the blue section outward would reduce $\|B_n\|_2$. A similar plot of the  The plots of the shape gradient of $\|B_n\|_2$ in Figures \ref{shape_grad_normB} and  \ref{shape_grad_normB_fixed} look qualitatively different, namely, the locations of high and low shape gradient have changed slightly and these peak values are larger with the $\|K\|_2$ constraint in place. The shape gradient appears to be greatest in the convex regions of the plasma surface with large curvature as seen in Figure \ref{shape_grad_normB} and Figure \ref{shape_grad_normK}. Pushing these regions inward would lower $\|B_n\|_2$ there, thus decreasing the magnetic field error. 

% This result is somewhat unexpected, as previous results \citep{paul2018} indicate that concave regions of the plasma surface require the winding surface to be closer to the plasma surface to maintain small field error. 
% We note that implementation of the constraint described in \S\ref{sec:future} may alter these results slightly.

\begin{figure}
    \centering
    \begin{subfigure}[b]{0.49\textwidth}
    \includegraphics[trim=4cm 3cm 3cm 3cm, clip,width=1.0\textwidth]{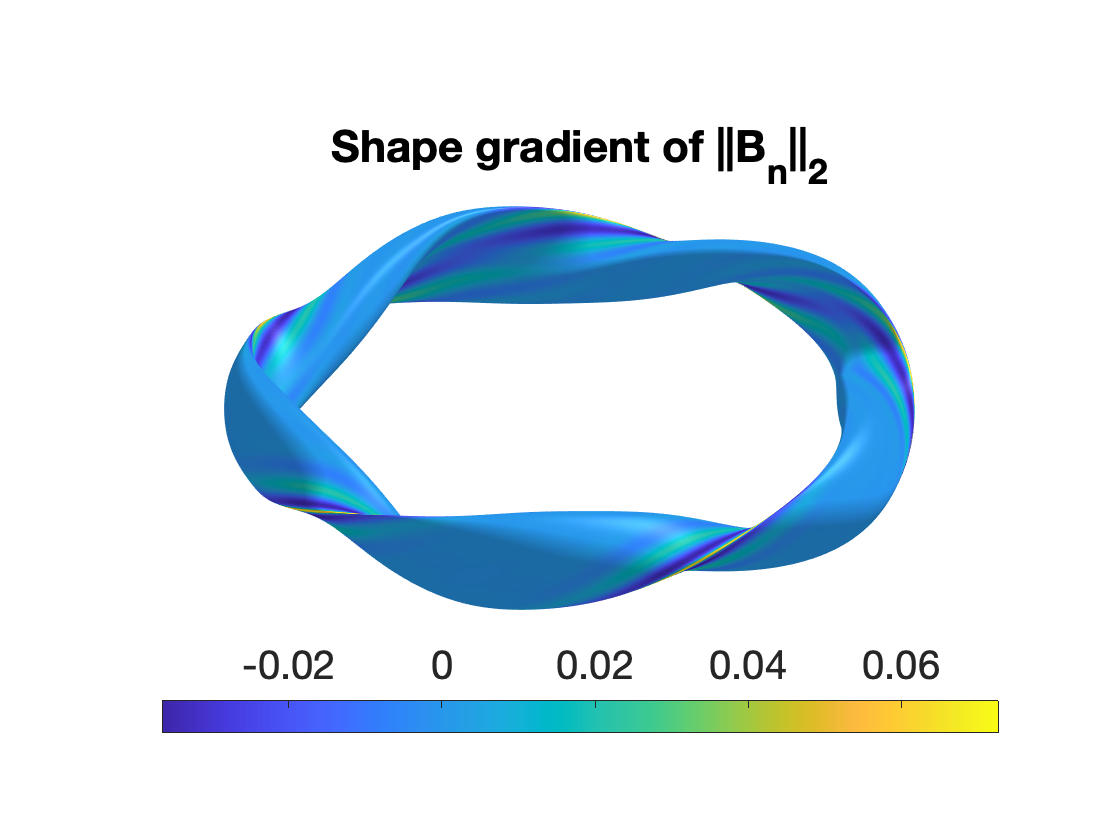}
    \caption{}
    \end{subfigure}
    \begin{subfigure}[b]{0.49\textwidth}
    \includegraphics[trim=4cm 3cm 3cm 3cm, clip,width=1.0\textwidth]{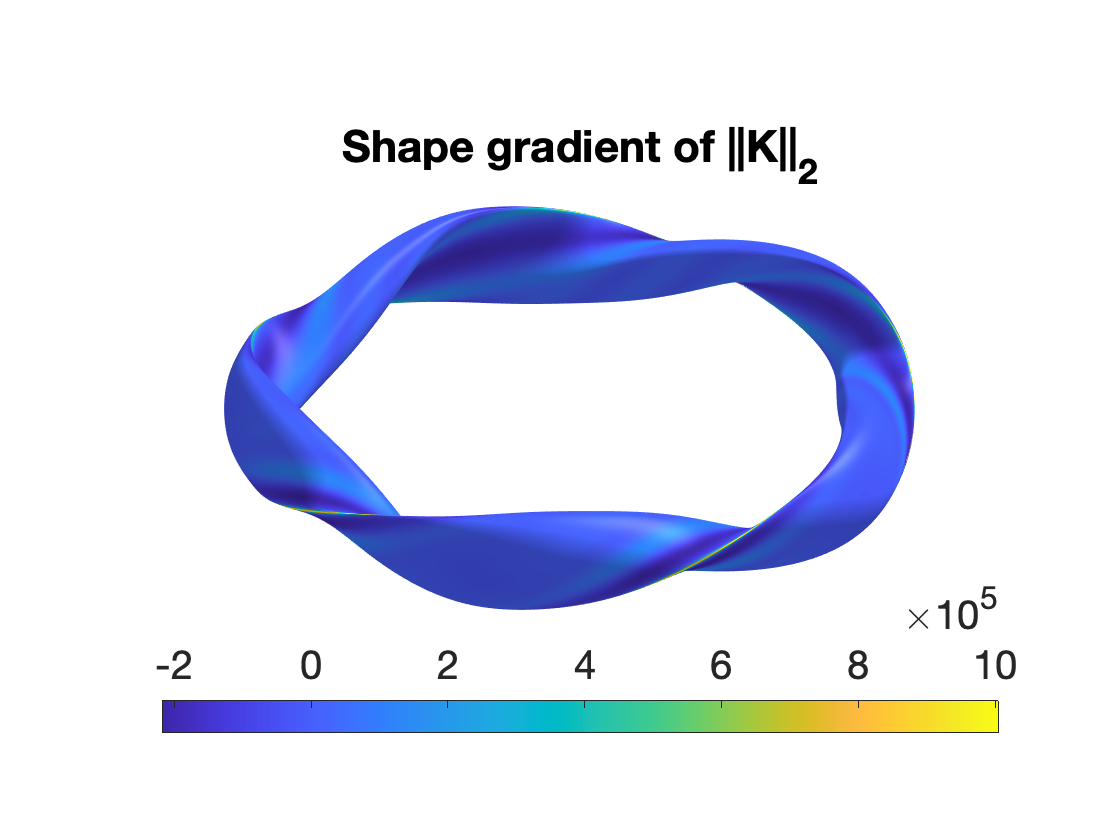}
    \caption{}
    \end{subfigure}
    \begin{subfigure}[b]{0.49\textwidth}
    \includegraphics[trim=4cm 3cm 3cm 3cm, clip,width=1.0\textwidth]{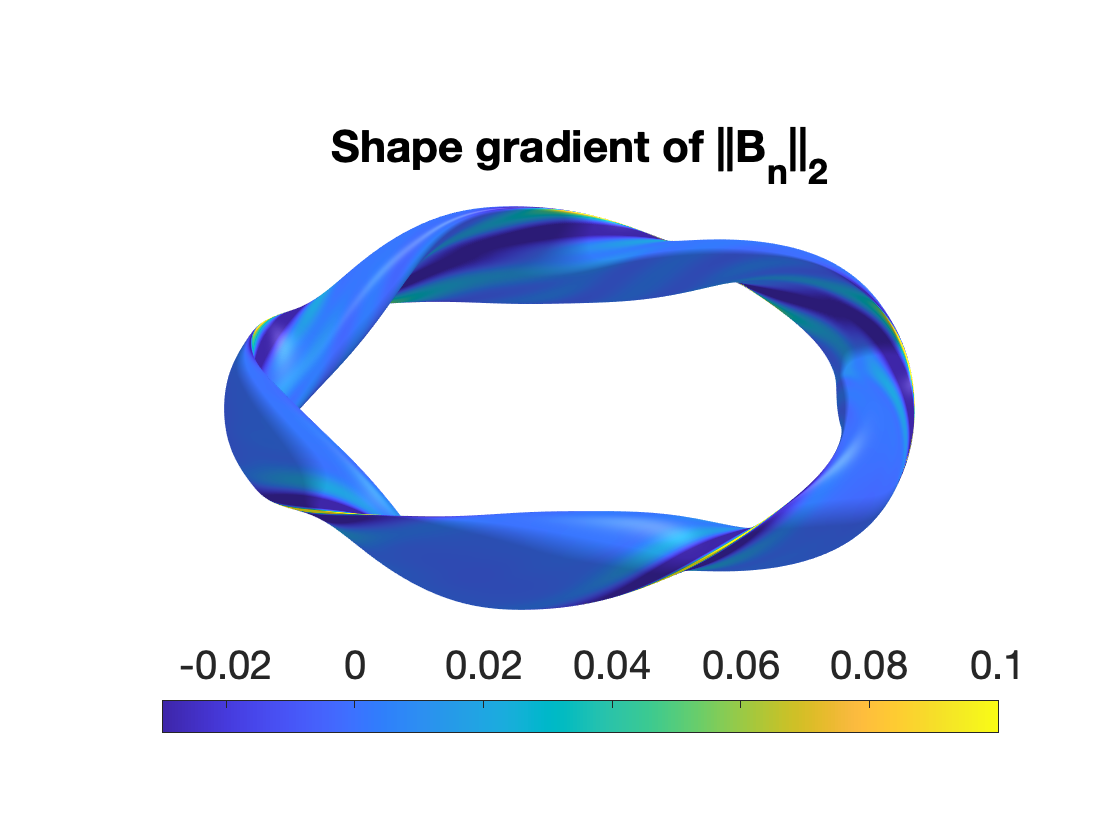}
    \caption{}
    \end{subfigure}    
    \begin{subfigure}[b]{0.49\textwidth}
    \includegraphics[trim=4cm 3cm 3cm 3cm, clip,width=1.0\textwidth]{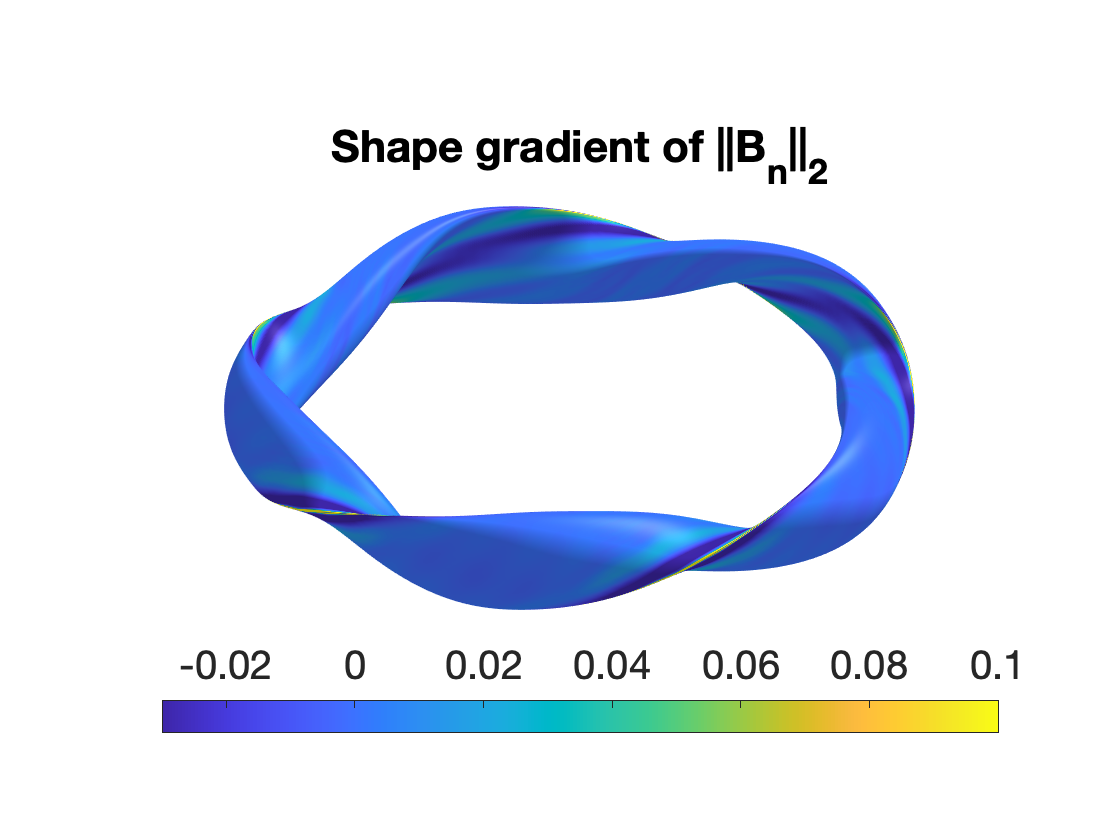}
    \caption{}
    \end{subfigure}    
    \caption{The shape gradient of (a) $\|B_n\|_2$, (b) $\|K\|_2$, (c) $\|B_n\|_2$ with fixed $\|K\|_2$, and (d) $\|B_n\|_2$ with fixed $\|K\|_2$ and fixed offset distance is computed for the W7-X boundary with a uniform offset winding surface of 0.37 m. The shape gradient is computed using a linear solve of (\ref{shape_grad_lin_sys}) for a square system with modes $m \le 30$ and $|n/N_P| \le 30$.}
    \label{fig:w7x_sg}
\end{figure}

In Figure \ref{fig:w7x_sg}, we present the results for the W7-X boundary.
% The shape gradient is computed with the same resolution {\color{red} and number of modes} used for the benchmark in Section \ref{sec:area_shape_gradient}. 
The parameter derivatives with respect to modes with $m \le 30$ and $|n/N_P| \le 30$ are computed on a grid with $N_{\overline{\theta}} = N_{\zeta} = 100$, and the shape gradient is discretized with a Fourier series with the same set of modes. We see that the shape gradient for $||B_n||_2$ is again peaked in the regions of large convexity and slightly negative in the regions of concavity. This indicates that the normal field could be reduced by pushing the surface inward in the sharply convex regions and outward in the concave regions. We see that the shape gradient for $||K||_2$ similarly has a large magnitude in the convex regions. Considering the shape gradient of $||B_n||_2$ with the additional constraints, we see that the trends are qualitatively similar.

Finally, we mention that the calculations presented in this Section rely on the choice of the offset distance, $a$. We find that the trends presented here do not strongly vary with this choice. This is consistent with the observations in \citep{Paul2018}.

\section{Conclusion}
We have presented a new approach to stellarator design which differs from the traditional approach of choosing a plasma shape for its MHD properties and then optimizing the coils to be consistent with the boundary. 
% Our approach puts the two steps on an equal footing by determining how to modify the plasma boundary in order to simplify the coil shapes. 
To assess a given plasma boundary, we evaluate metrics obtained from the current potential on a uniformly offset winding surface, namely the normal field error and current density. While similar metrics have been included in fixed-boundary optimization with the ROSE code \citep{drevlak}, the derivatives of these metrics with respect to the plasma boundary have not previously been computed. To enable gradient-based optimization, we compute derivatives of these objectives with respect to the Fourier series coefficients used to parameterize the plasma surface. An adjoint method is used in order to reduce the number of linear solves required to evaluate the parameter derivatives. Moreover, we present a new parameterization of the plasma boundary which improves the convergence of the Fourier series by choosing a fixed poloidal angle to be an arclength angle on a uniformly offset surface. This choice, furthermore, eliminates the non-uniqueness of the poloidal angle that is present in the standard cylindrical representation. If the plasma boundary encloses a star domain, the related null space in the optimization of the boundary can be eliminated. While spectral condensation is not novel for stellarator calculations, computing our proposed poloidal angle does not require nonlinear optimization as the Hirshman-Breslau method does \citep{Hirshman1998}.
% The poloidal angle is uniquely defined using an arclength angle on the plasma boundary or a uniformly offset surface. 

With these parameter derivatives, 
% $\chi^2_B$ and $\chi^2_K$ as well as their normalized quantities $\|B_n\|_2$ and $\|K\|_2$ with respect to the Fourier series coefficients used to parameterize the plasma surface, 
we construct the shape gradient with respect to displacements of the plasma boundary. This provides information on how to deform the plasma surface in order to reduce the normal field error or coil complexity. Several constraints are enforced in order to ensure that the regularization parameter used in the REGCOIL code is chosen to target engineering constraints and that the winding surface maintains a uniform displacement from the plasma surface. While the constraint on the winding surface amplifies high-frequency noise in the plasma boundary, as higher-order derivatives of the position vector are required, such noise can be reduced with the application of our proposed poloidal angle. The resulting shape gradient provides intuition about what kinds of plasma shapes are consistent with simpler coils. The results presented in Section \ref{sec:shape_gradient_metrics} indicate that both convex and concave curvature of the plasma boundary are associated with coil complexity. The correlation between concavity of the plasma boundary and coil complexity has been noted in the literature \citep{landreman2016,Paul2018}. Furthermore, we note that several coil design studies have indicated increased normal field error near convex regions \citep{Strickler2003,Zhu2018}. We remark that convex surface shapes are sometimes easy to produce with external coils, such as the x-point of a diverted tokamak. Further analysis of such cases is necessary to distinguish if these trends are captured by our shape gradient method.

The shape gradient obtained with this method could be used within the optimization of an MHD equilibrium to avoid arriving at a configuration that requires overly-complex coils. While the parameter space is taken to be the shape of the plasma boundary during fixed-boundary optimization, the coil complexity metrics can be computed from the current potential solution on a uniformly offset winding surface. Using the shape gradient of the normal field error in the gradient-based optimization of the equilibrium would allow for the plasma surface to be more accurately reconstructed by REGCOIL while achieving a desired coil complexity. This optimization would need to be balanced with the optimization of the MHD properties of the equilibrium. With derivatives computed efficiently from the adjoint method, gradient-based optimization of the equilibrium could be performed. With a slight modification, a similar technique could be utilized to compute the shape gradient of objectives related to the permanent magnets \citep{Helander2020,Zhu2020,Landreman2020} required to confine a given equilibrium. Under the assumption that a continuous surface magnetization lies on the winding surface with an orientation in the normal direction, an equivalent linear least-squares problem can be formulated in which the current potential is related to the magnitude of the magnetization. Thus, in computing the shape gradient with respect to the plasma boundary, the norm of the current potential rather than the norm of the current density could be fixed to choose the regularization parameter.

The new parameterization of the plasma boundary utilizes the distance from a fixed coordinate axis. In this work, we have taken this coordinate axis to coincide with the magnetic axis, although this assumption is not required. Other possible choices for the coordinate axis may be favorable. In the stellarator equilibrium code VMEC, the initial coordinate axis is taken to maximize the minimum value of the flux coordinate Jacobian \citep{hirshman}, and in the SPEC code \citep{Hudson2011} the coordinate axis is chosen to minimize the variation of the Jacobian over a poloidal cross-section \citep{Qu2020}. Although a three-dimensional Jacobian is not necessary for current potential calculations, a similar analysis could be performed to minimize the variation of the surface Jacobian ($|\partial \bm{r}/\partial \overline{\theta} \times \partial \bm{r}/\partial \zeta|$) with a cleverly chosen axis.

% These give information on how to deform the plasma surface to alter the values of $\|B_n\|_2$ or $\|K\|_2$. We then consider a constraint on the normalized quantity $\|K\|_2$ calculated from $\chi^2_K$ to fix its value. Since the coil complexity is correlated with $\|K\|_2$, this fixes the coil complexity. With the derivatives with respect to the Fourier coefficients altered for this constraint, the shape gradient of $\|B_n\|_2$ tells us how to deform the plasma surface to alter $\|B_n\|_2$ while maintaining a desired coil complexity. 

\section*{Funding}
This work was supported by the US Department of Energy FES grants (E.J.P., grant numbers DE-FG02-93ER-54197, DE-FC02-08ER-54964); and the ARCS Foundation (E.J.P.). Some computations presented in this work have used resources at the National Energy Research Scientific Computing Center (NERSC).

\section*{Declaration of interests}
The authors report no conflict of interest.

\section*{Acknowledgements}
We would like to thank M. Landreman for use of the REGCOIL code and for discussion of a uniformly offset coil constraint. We would also like to thank T. Antonsen and M. Landreman for providing feedback on a draft of this manuscript. 
%This work was supported by the US Department of Energy FES grants DE-FG02-93ER-54197 and DE-FC02-08ER-54964 and the ARCS Foundation. Some computations presented in this work have used resources at the National Energy Research Scientific Computing Center (NERSC).

\appendix
\section{Details of adjoint calculations}

\subsection{Fixed Norm Constraint}
\label{app:fixed_norm}

To obtain $\p \boldsymbol{\Phi} / \p \Omega_j$ subject to the constraints $\boldsymbol{F}=0$ and $G=0$, the differential of $\lambda$ must be eliminated from (\ref{eq:constraint_differentials}). First, we solve $\mathrm{d}\boldsymbol{F} = 0$ for $\mathrm{d}\boldsymbol{\Phi}$ and substitute this into $\mathrm{d}G = 0$ to obtain,
\begin{equation}
\begin{split}
    \mathrm{d}G \big|_{\boldsymbol{F}=0} &= \sum_j \frac{\p G}{\p \Omega_j}\Bigg|_{\mathrm{d}\boldsymbol{\Phi}=0} \mathrm{d}\Omega_j\\
    &- \frac{\p G}{\p \boldsymbol{\Phi}} \cdot \mathsfbi{A}^{-1} \left[ \sum_j \left( \frac{\p \mathsfbi{A}}{\p \Omega_j}\boldsymbol{\Phi} - \frac{\p \boldsymbol{b}}{\p \Omega_j} \right)\mathrm{d}\Omega_j +\left( \mathsfbi{A}^K \boldsymbol{\Phi} - \boldsymbol{b}^K \right)\mathrm{d}\lambda \right] = 0.
\end{split}
\end{equation}
The term with the dot product has a similar form to (\ref{eq:dot_to_transpose}), so we define an additional adjoint variable $\widetilde{\bm{q}}$ which satisfies $\mathsfbi{A}^T \widetilde{\boldsymbol{q}}=\p G / \p \boldsymbol{\Phi}$, (\ref{eq:fixed_norm_adjoint}).
% \begin{equation}
%     \mathsfbi{A}^T \widetilde{\boldsymbol{q}}=\frac{\p G}{\p \boldsymbol{\Phi}}.
%     \label{eq:fixed_norm_adjoint}
% \end{equation}
The use of this adjoint variable, once again, avoids having to solve a linear system for each $\Omega_j$. The equation $\mathrm{d}G = 0$ at fixed $\boldsymbol{F}$ can then be solved for $\mathrm{d}\lambda$ subject to the constraints $\boldsymbol{F}=0$ and $G=0$,
\begin{equation}
    \mathrm{d}\lambda \big|_{\boldsymbol{F}=0,\ G=0} = \frac{\sum_j \left[ \frac{\p G}{\p \Omega_j}\Big|_{\mathrm{d}\boldsymbol{\Phi}=0} - \widetilde{\boldsymbol{q}}\cdot \left( \frac{\p \mathsfbi{A}}{\p \Omega_j}\boldsymbol{\Phi} - \frac{\p \boldsymbol{b}}{\p \Omega_j} \right) \right]\mathrm{d}\Omega_j}{\widetilde{\boldsymbol{q}}\bcdot(\mathsfbi{A}^K\boldsymbol{\Phi} - \boldsymbol{b}^K)} .
\end{equation}
The dependence on $\lambda$ in (\ref{eq:constraint_differentials}) can now be eliminated to obtain the differential of $\boldsymbol{\Phi}$ subject to both the constraints,
\begin{equation}
\begin{split}
    \mathrm{d}\boldsymbol{\Phi} \big|_{\boldsymbol{F}=0,\ G=0} &= \sum_j -\mathsfbi{A}^{-1}\Bigg[ \left( \frac{\p \mathsfbi{A}}{\p \Omega_j}\boldsymbol{\Phi} - \frac{\p \boldsymbol{b}}{\p \Omega_j} \right)\\
    &+ \frac{(\mathsfbi{A}^K\boldsymbol{\Phi} - \boldsymbol{b}^K)}{\widetilde{\boldsymbol{q}}\bcdot(\mathsfbi{A}^K\boldsymbol{\Phi} - \boldsymbol{b}^K)} \left[ \frac{\p G}{\p \Omega_j}\Bigg|_{\mathrm{d}\boldsymbol{\Phi}=0} - \widetilde{\boldsymbol{q}}\cdot \left( \frac{\p \mathsfbi{A}}{\p \Omega_j}\boldsymbol{\Phi} - \frac{\p \boldsymbol{b}}{\p \Omega_j} \right) \right] \Bigg]\mathrm{d}\Omega_j.
\end{split}
\end{equation}
Since this is now of the form of the chain rule, we conclude (\ref{eq:fixed_norm_implicit}).

\subsection{Fixed Offset Constraint}
\label{app:fixed_offset}
% This integrand is related to the derivative of the coil normal vector by,
% \begin{equation}
%     \frac{\p N'}{\p \Omega_j} = \frac{\p \boldsymbol{N}'}{\p \Omega_j} \bcdot \boldsymbol{\hat{n}}'.
% \end{equation}
% The coil normal vector is given by,
% \begin{equation}
%     \boldsymbol{N}' = \frac{\p \boldsymbol{r}'}{\p \zeta'} \times \frac{\p \boldsymbol{r}'}{\p \theta'},
% \end{equation}
% so its derivative with respect to the plasma surface parameters is,
% \begin{equation}
%     \frac{\p \boldsymbol{N}'}{\p \Omega_j} = \frac{\p^2 \boldsymbol{r}'}{\p \zeta' \p \Omega_j} \times \frac{\p \boldsymbol{r}'}{\p \theta'} + \frac{\p \boldsymbol{r}'}{\p \zeta'} \times \frac{\p^2 \boldsymbol{r}'}{\p \theta' \p \Omega_j},
% \end{equation}
% and similarly for the plasma normal vector. 

The outward normal vector for a surface $\boldsymbol{r}$ parameterized by two angles $\theta$ and $\zeta$ is given by,
\begin{equation}
    \boldsymbol{N} = \frac{\p \boldsymbol{r}}{\p \zeta} \times \frac{\p \boldsymbol{r}}{\p \theta},
\end{equation}
and the derivative of the magnitude of this vector with respect to a quantity $Q$ is given by,
\begin{equation}
    \frac{\p N}{\p Q} = \frac{\p \boldsymbol{N}}{\p Q} \bcdot \boldsymbol{\hat{n}},
\end{equation}
where $\boldsymbol{\hat{n}}$ is the vector in the direction of $\boldsymbol{N}$ with unit length. These equations apply to both the plasma and coil winding surfaces. For the constraint (\ref{eq:coil_offset}), the derivative of the coil winding surface position vector with respect to some quantity $Q_1$, on which the plasma surface position vector depends is given by,
\begin{equation}
    \frac{\p \boldsymbol{r}'}{\p Q_1} = \frac{\p \boldsymbol{r}}{\p Q_1} + \frac{a}{N}\left( \frac{\p \boldsymbol{N}}{\p Q_1} - \frac{\p N}{\p Q_1} \boldsymbol{\hat{n}} \right),
    \label{eq:cp_coupling_1}
\end{equation}
where $\boldsymbol{N}$ is the outward normal vector to the plasma surface, and $N$ is the magnitude of this vector. Furthermore, differentiation with respect to a second quantity $Q_2$ gives,
\begin{equation}
\begin{split}
    \frac{\p^2 \boldsymbol{r}'}{\p Q_1 \p Q_2} &= \frac{\p^2 \boldsymbol{r}}{\p Q_1 \p Q_2} + \frac{a}{N^2}\Bigg[ N\frac{\p^2 \boldsymbol{N}}{\p Q_1 \p Q_2} - \frac{\p^2 N}{\p Q_1 \p Q_2} \boldsymbol{N}\\
    &- \left( \frac{\p N}{\p Q_1}\frac{\p \boldsymbol{N}}{\p Q_2} + \frac{\p N}{\p Q_2}\frac{\p \boldsymbol{N}}{\p Q_1} \right) + 2\frac{\p N}{\p Q_1}\frac{\p N}{\p Q_2} \boldsymbol{\hat{n}} \Bigg].
\end{split}
\label{eq:cp_coupling_2}
\end{equation}
The derivatives of the coil position vector with respect to the angles, which are the same for the plasma and coil surfaces with this constraint, are given by (\ref{eq:cp_coupling_1}), and the second derivatives of the coil position vector with respect to the angles and the plasma parameters are given by (\ref{eq:cp_coupling_2}). In this way, the plasma normal vector and area depend on the first derivatives of the plasma position with respect to the angles, while the coil normal vector and area also depend on the second derivatives of the plasma position with respect to the angles. 

\subsection{Fixed Norm and Offset Constraints}
\label{app:fixed_norm_and_offset}
For the case of (\ref{fixed_L2K}) and the fixed offset constraint, the explicit dependence of $G$ on $\Omega_j$ is given by,
\begin{equation}
    \frac{\p G }{\p \Omega_j}\Bigg|_{\mathrm{d}\boldsymbol{\Phi}=0} = \frac{1}{2\|\boldsymbol{K}\|_2 a_{\text{coil}}} \left( \frac{\p \chi^2_K}{\p \Omega_j}\Bigg|_{\mathrm{d}\boldsymbol{\Phi}=0} - \frac{\chi^2_K}{a_{\text{coil}}} \frac{\p a_{\text{coil}}}{\p \Omega_j} \right),
\end{equation}
and,
\begin{equation}
    \frac{\p G }{\p \boldsymbol{\Phi}} = \frac{1}{2\|\boldsymbol{K}\|_2 a_{\text{coil}}} \frac{\p \chi^2_K}{\p \boldsymbol{\Phi}}.
\end{equation}
This means that we also have,
\begin{equation}
    \widetilde{\boldsymbol{q}} = \frac{1}{2\|\boldsymbol{K}\|_2 a_{\text{coil}}} \boldsymbol{q}_K,
\end{equation}
and $\boldsymbol{q}_B$ is also proportional to $\widetilde{\boldsymbol{q}}$ due to \eqref{eq:adjoint_variable_proportional}.
Note the appearance of the same proportionality constant between these adjoint variables in the expression for $\p G /\p \Omega_j$. When these are substituted into (\ref{eq:fixed_norm_implicit}), this proportionality factor cancels, leaving,
\begin{equation}
\begin{split}
    \frac{\p \boldsymbol{\Phi} }{\p \Omega_j}&\Bigg|_{\boldsymbol{F}=0,\ G=0, \ \boldsymbol{r}'=\boldsymbol{r}+a\boldsymbol{\hat{n}}} = \\ -\mathsfbi{A}^{-1}\Bigg[
    &\left( \frac{\p \mathsfbi{A}}{\p \Omega_j}\boldsymbol{\Phi} - \frac{\p \boldsymbol{b}}{\p \Omega_j} \right) - \frac{(\mathsfbi{A}^K\boldsymbol{\Phi} - \boldsymbol{b}^K)}{\boldsymbol{q}_K \bcdot(\mathsfbi{A}^K\boldsymbol{\Phi} - \boldsymbol{b}^K)} \left( \boldsymbol{q}_K \cdot \left( \frac{\p \mathsfbi{A}}{\p \Omega_j}\boldsymbol{\Phi} - \frac{\p \boldsymbol{b}}{\p \Omega_j} \right) \right)\\
    +& \frac{(\mathsfbi{A}^K\boldsymbol{\Phi} - \boldsymbol{b}^K)}{\boldsymbol{q}_K \bcdot(\mathsfbi{A}^K\boldsymbol{\Phi} - \boldsymbol{b}^K)} \left( \frac{\p \chi^2_K}{\p \Omega_j}\Bigg|_{\mathrm{d}\boldsymbol{\Phi}=0} - \frac{\chi^2_K}{a_{\text{coil}}} \frac{\p a_{\text{coil}}}{\p \Omega_j} \right) \Bigg].
\end{split}
\end{equation}
When this is used in (\ref{tot_chi2_dep}), the leading $\mathsfbi{A}^{-1}$ will be moved to the other side of the dot product so that the previous adjoint variables, $\boldsymbol{q}_B$ and $\boldsymbol{q}_K$, are dotted with the expression in brackets. For $\chi^2_K$, the first line of the bracketed expression vanishes, leaving,
\begin{equation}
    \frac{\p \chi^2_K}{\p \boldsymbol{\Phi}}\bcdot\\
    \frac{\p \boldsymbol{\Phi}}{\p \Omega_j} = -\frac{\p \chi^2_K}{\p \Omega_j}\Bigg|_{\mathrm{d}\boldsymbol{\Phi}=0} + \frac{\chi^2_K}{a_{\text{coil}}} \frac{\p a_{\text{coil}}}{\p \Omega_j}.
\end{equation}
For $\chi^2_B$, we note that $\p \chi^2_B / \p \boldsymbol{\Phi}$ and $\p \chi^2_K / \p \boldsymbol{\Phi}$ are related by (\ref{eq:BK_relation}); hence,
\begin{equation}
    \frac{\p \chi^2_B}{\p \boldsymbol{\Phi}}\bcdot\\
    \frac{\p \boldsymbol{\Phi}}{\p \Omega_j} = -\lambda \frac{\p \chi^2_K}{\p \boldsymbol{\Phi}}\bcdot\\
    \frac{\p \boldsymbol{\Phi}}{\p \Omega_j} = -\lambda \left(\frac{\chi^2_K}{a_{\text{coil}}} \frac{\p a_{\text{coil}}}{\p \Omega_j} - \frac{\p \chi^2_K}{\p \Omega_j}\Bigg|_{\mathrm{d}\boldsymbol{\Phi}=0} \right).
\end{equation}
For the constraint (\ref{fixed_L2K}) without the fixed offset constraint, both $a_{\text{coil}}$ and $\chi^2_K$ have no explicit dependence on $\Omega_j$. It can be seen from these equations that for this case, the implicit dependencies of $\chi^2_B$ and $\chi^2_K$ vanish; thus, their total derivatives reduce to only the explicit dependencies on the plasma parameters.

\bibliographystyle{jpp}

\bibliography{Sources}

\end{document}